\documentclass[envcountsame]{llncs}

\makeatletter
\RequirePackage[bookmarks,unicode,colorlinks=true]{hyperref}%
   \def\@citecolor{blue}%
   \def\@urlcolor{blue}%
   \def\@linkcolor{blue}%

\def\orcidID#1{\smash{\href{http://orcid.org/#1}{\protect\raisebox{-1.25pt}{\protect\includegraphics{orcid_color.eps}}}}}
\makeatother

\usepackage{xcolor}
\usepackage{mathtools}
\usepackage{multirow}
\usepackage{stmaryrd}
\usepackage{cancel}
\usepackage{amsmath,amssymb}
\usepackage{thmtools, thm-restate}
\usepackage[shortlabels]{enumitem}
\usepackage{microtype}
\usepackage{pbox}
\usepackage{marvosym}
\usepackage[ruled, vlined, linesnumbered]{algorithm2e}
\hypersetup{
	pdftitle={Automatic Complexity Analysis of Integer Programs via Triangular Weakly Non-Linear Loops}, colorlinks=true, linkcolor=blue, citecolor=olive, filecolor=magenta, urlcolor=cyan
}
\usepackage{todonotes}
\usepackage{placeins}
\usepackage{tikz}
\usepackage{caption}
\usepackage{subcaption}
\usepackage{mymatrix}

\usetikzlibrary{shapes,calc,arrows,automata,decorations.pathmorphing}
\usepackage[capitalize,nameinlink]{cleveref}
\usepackage{IEEEtrantools}

\newenvironment{myproof}{
	\noindent{\it Proof.}
}{\qed
	\medskip
}
\makeatletter \newcommand*\bigcdot{\mathpalette\bigcdot@{.4}}
\newcommand*\bigcdot@[2]{\mathbin{\vcenter{\hbox{\scalebox{#2}{$\m@th#1\bullet$}}}}}
\makeatother

\renewcommand{\emptyset}{\varnothing}

\newcommand{\tool}[1]{\textsf{#1}}
\newcommand{\emphit}[1]{\textit{#1}}

\newcommand{\indv}{d}

\newcommand{\dontprint}[1]{}
\newcommand{\xvec}{\vec{x}}

\DeclarePairedDelimiter\abs{\lvert}{\rvert}

\newcommand{\NN}{\mathbb{N}}
\newcommand{\ZZ}{\mathbb{Z}}

\newcommand{\NNC}{\overline{\mathbb{N}}}
\newcommand{\QQ}{\mathbb{Q}}

\newcommand{\PP}{\mathcal{P}}

\newcommand{\MRF}{\text{M}\Phi\text{RF}}
\newcommand{\MRFs}{\text{M}\Phi\text{RFs}}

\newcommand{\NNPPEE}{\mathbb{NPE}}

\newcommand{\var}{\texttt}

\newcommand{\true}{\var{true}}
\newcommand{\false}{\var{false}}

\newcommand{\cl}[1]{\texttt{cl}^n_{#1}}
\newcommand{\clExp}[2]{\texttt{cl}^{#2}_{#1}}

\newcommand{\ud}{a_1}
\newcommand{\ub}{b_1}
\newcommand{\ld}{a_2}
\newcommand{\lb}{b_2}

\newcommand{\lex}{>_{\mathrm{lex}}}

\newcommand{\location}{\ell}
\newcommand{\valuation}{\sigma}
\newcommand{\initial}{\sigma_0}
\newcommand{\Valuation}{\Sigma}
\newcommand{\update}{\eta}

\newcommand{\guard}{\varphi}

\newcommand{\landau}{\mathcal{O}}

\newcommand{\entry}{\mathcal{E}}

\newcommand{\Size}{{\mathcal{SB}}}

\newcommand{\loc}{{\mathcal{RB}_{\text{loc}}}}
\newcommand{\glo}{{\mathcal{RB}_{\text{glo}}}}
\newcommand{\glopr}{{\mathcal{RB}'_{\text{glo}}}}
\DeclareMathOperator{\rc}{rc}

\newcommand{\pret}{r}
\newcommand{\actt}{t}
\newcommand{\prestate}{\valuation}
\newcommand{\actstate}{{\tilde{\valuation}}}
\newcommand{\actl}{{\tilde{\location}}}
\newcommand{\prel}{\location}

\newcommand{\BoundSet}{\mathcal{B}}

\newcommand{\AtomSet}{\mathcal{A}}

\newcommand{\FormulaSet}{\mathcal{F}}

\newcommand{\TSet}{\mathcal{T}}
\newcommand{\VSet}{\mathcal{V}}

\newcommand{\PVSet}{\mathcal{PV}}

\newcommand{\LSet}{\mathcal{L}}

\newcommand{\invariant}{\psi}
\newcommand{\InvSet}{\Psi}

\newcommand{\IntLoop}{(\invariant, \guard, \update)}
\newcommand{\IntProgram}{(\PVSet,\LSet,\location_0,\TSet)}

\newcommand{\braced}[1]{\lbrace #1 \rbrace}

\newcommand{\timeboundterm}{
	\sup \braced{ k \in \NN \mid \exists \, (\location', \valuation').\; (\location_0, \valuation_0) \; (\rightarrow^*_{\TSet} \circ \rightarrow_t)^k \; (\location', \valuation') }
}

\newcommand{\sizeboundterm}{
	\braced{ |\valuation'(v)| \mid \exists\, (\location', \valuation'). \; (\location_0, \valuation_0) \; (\rightarrow^* \circ \rightarrow_\actt) \; (\location', \valuation')}
}

\newcommand{\chain}{\star}

\crefname{definition}{Def.}{Def.}
\crefname{example}{Ex.}{Ex.}
\crefname{counterexample}{Counterex.}{Counterex.}
\crefname{appendix}{App.}{App.}
\crefname{ex}{Ex.}{Ex.}
\crefname{theorem}{Thm.}{Thm.}
\crefname{lemma}{Lemma}{Lemmas}
\crefname{remark}{Rem.}{Rem.}
\crefname{section}{Sect.}{Sect.}
\crefname{subsection}{Sect.}{Sect.}
\crefname{subsubsection}{Sect.}{Sect.}
\crefname{line}{Line}{Lines}
\crefname{corollary}{Cor.}{Cor.}
\crefname{figure}{Fig.}{Fig.}
\crefname{enumi}{}{}
\crefname{algorithm}{Alg.}{Alg.}

\newcommand{\kernel}[3]{\lceil#1\rceil^{#2}_{#3}}

\newcommand{\KoAT}[0]{\tool{KoAT}}

\newcommand{\sth}{\operatorname{sth}}
\newcommand{\sthb}{{\operatorname{sth}^{\sqcup}}}
\newcommand{\sthbInv}[1]{{\operatorname{sth}^{\sqcup}_{#1}}}
\DeclareMathOperator{\sign}{sign}
\SetKwInOut{Input}{input}
\SetKwInOut{Output}{output}
\newcommand*{\return}{\textbf{return}\,\,}

 \usepackage[location=appendix,prependtoappendix=true]{moveproofs}
 \newcommand{\paper}[1]{}
 \newcommand{\report}[1]{#1}

 \report{
  \setlength{\textwidth}{125mm}
   \setlength{\textheight}{198mm}
 }
 
\paper{\usepackage[ hyperref=true, backend=bibtex, firstinits=true, maxbibnames=99, sortcites, style=numeric-comp ]{biblatex}
\addbibresource{KoAT-TWN.bib}}
\report{\usepackage{cite}}

\title{\hspace*{-.45cm}\mbox{Automatic Complexity Analysis of Integer Programs} via Triangular Weakly Non-Linear Loops\thanks{funded by the Deutsche Forschungsgemeinschaft (DFG, German Research Foundation) - 235950644 (Project GI 274/6-2) and DFG Research Training Group 2236 UnRAVeL}}
\author{Nils Lommen\paper{\orcidID{0000-0003-3187-9217}} \and Fabian Meyer\paper{\orcidID{0000-0003-1038-4944}} \and Jürgen Giesl\paper{\orcidID{0000-0003-0283-8520}}}
\institute{LuFG Informatik 2, RWTH Aachen University, Germany}

\begin{document}
\allowdisplaybreaks

\maketitle \begin{abstract}
	There exist several results on deciding termination and computing runtime bounds for \emph{triangular weakly non-linear loops} (twn-loops).
        We show how to use results on such 
        subclasses of programs where complexity bounds are computable within incomplete
        approaches for complexity analysis of full integer programs. To this end, we present a novel modular approach which computes local runtime bounds for subprograms which can be transformed into twn-loops.
	These local runtime bounds are then lifted to global runtime bounds for the whole program.
The power of our\linebreak approach is shown by our implementation in the tool
$\tool{KoAT}$ which analyzes complexity of programs where all other state-of-the-art tools fail.
\end{abstract}

\section{Introduction}\label{sec-introduction}
Most approaches for automated complexity analysis of programs are
based on incomplete techniques like ranking functions (see,
e.g.,  \cite{albert2008AutomaticInferenceUpper,flores-montoya2014ResourceAnalysisComplex,heizmann2015RankingTemplatesLinear,albert2019ResourceAnalysisDriven,brockschmidt2016AnalyzingRuntimeSize,ben-amram2017MultiphaseLinearRankingFunctions,ramlpopl17,Festschrift,ben-amram2019MultiphaseLinearRankingFunctions,ben-amram2017MultiphaseLinearRankingFunctions,sinn2017ComplexityResourceBound,cofloco2}).
However, there also exist numerous results on subclasses of programs where questions concerning
termination or complexity are \emph{decidable}, e.g.,
\cite{dblp:conf/cav/braverman06,dblp:conf/icalp/hosseinio019,dblp:conf/cav/tiwari04,hark2020PolynomialLoopsTermination,frohn2019TerminationTriangularInteger,kincaid19,kovacs08,frohn2020TerminationPolynomialLoops,xu13}.
In this work we consider the subclass of \emph{triangular weakly non-linear loops}
(twn-loops), where there exist \emph{complete} techniques for analyzing termination and runtime complexity
(we discuss the ``completeness'' and decidability of these techniques below).
An example for a twn-loop is:
\begin{equation}
	\label{WhileExample}
	\textbf{while } (x_1^2 + x_3^5 < x_2 \, \wedge \, x_1 \neq 0) \textbf{ do }
(x_1, x_2, x_3) \leftarrow (-2\cdot x_1, \,	3\cdot x_2 - 2\cdot x_3^3,  \, x_3) \quad
      \end{equation}
Its guard is a propositional formula over (possibly \emphit{non-linear}) polynomial inequations.
The update is \emph{weakly non-linear}, i.e., no variable $x_i$ occurs
non-linear in its own update.
Furthermore, it is \emph{triangular}, i.e., we can order the variables such that
the update of any $x_i$ does not depend on the variables $x_1, \ldots, x_{i-1}$ with
smaller indices.
Then, by handling one variable after the other one can compute a \emph{closed form} which
corresponds to applying the loop's update $n$ times.
Using these closed\linebreak forms, termination can be reduced to an existential formula over
$\ZZ$ \cite{frohn2020TerminationPolynomialLoops} (whose validity is decidable for linear arithmetic and where SMT solvers
often also prove (in)validity in the non-linear case).
In this way, one can show that non-termination of twn-loops over $\ZZ$ is semi-decidable
(and it is decidable over the real numbers).

While 
termination of
twn-loops over $\ZZ$ is not decidable, by  using the closed
forms,  \cite{hark2020PolynomialLoopsTermination}
presented a ``\emph{complete}''
complexity analysis technique. More precisely,
for every twn-loop  over $\ZZ$, it infers a polynomial
which is an upper bound on the runtime for all those inputs where the loop terminates.
So for all (possibly non-linear) terminating twn-loops over $\ZZ$, the technique of
\cite{hark2020PolynomialLoopsTermination}  \emph{always} computes polynomial runtime bounds. 
In contrast, existing tools based on incomplete techniques for complexity analysis often
fail for programs with non-linear arithmetic.

In \cite{brockschmidt2016AnalyzingRuntimeSize,Festschrift} we presented
such an incomplete 
modular technique for
complexity analysis which uses individual ranking functions for different
subprograms.
Based on this, we now introduce a novel
approach to automatically infer runtime bounds\linebreak for programs possibly consisting of
multiple consecutive or nested loops by
han\-dling some subprograms as twn-loops and by using ranking functions for others.
In order to compute runtime bounds, we analyze subprograms in topological order, i.e., in
case of multiple consecutive loops, we start with the first loop and propagate knowledge
about the resulting values of variables to subsequent loops.
By inferring runtime bounds for one subprogram after the other, in the end we obtain
a bound on the runtime complexity of the whole program.
We first try to compute runtime bounds for subprograms by so-called multiphase linear ranking functions ($\MRFs$, see
 \cite{Festschrift,ben-amram2017MultiphaseLinearRankingFunctions,heizmann2015RankingTemplatesLinear,ben-amram2019MultiphaseLinearRankingFunctions}).
If $\MRFs$ do not yield a finite runtime bound for the respective subprogram, then we use our novel twn-technique on the unsolved parts of the subprogram.
So for the first time, ``complete''
complexity analysis
techniques like \cite{hark2020PolynomialLoopsTermination} for subclasses of programs with
\emph{non-linear} arithmetic are combined with incomplete techniques based on (linear) ranking functions like \cite{brockschmidt2016AnalyzingRuntimeSize,Festschrift}.
Based on our approach, in future work one could  integrate ``complete'' techniques for further subclasses
(e.g., for \emph{solvable loops}
\cite{solvable-maps,kincaid19,xu13,kovacs08}
which can be transformed
into twn-loops by suitable automorphisms  \cite{frohn2020TerminationPolynomialLoops}).

\paragraph{Structure: }
After introducing preliminaries in \cref{Preliminaries}, in \cref{Computing Global Runtime
  Bounds} we show how to lift a (local) runtime bound which is only sound for a subprogram
to an overall global runtime bound.
In contrast to previous techniques
\cite{brockschmidt2016AnalyzingRuntimeSize,Festschrift}, our lifting approach works for
any method of bound computation (not only for ranking functions).
In \cref{Local Runtime Bounds for TWN-Self-Loops}, we improve the existing results on
complexity analysis of twn-\linebreak loops \cite{frohn2019TerminationTriangularInteger,hark2020PolynomialLoopsTermination,frohn2020TerminationPolynomialLoops}
such that they yield concrete polynomial bounds, we refine these bounds by
considering invariants,
and we show how to apply these results to full programs which contain twn-loops as subprograms.
\cref{Local Runtime Bounds for TWN-Cycles} extends this technique to larger subprograms which can be transformed into twn-loops.
In \cref{Evaluation} we evaluate the implementation of our approach in the
complexity analysis tool \tool{KoAT}
and show that one can now also
successfully analyze the runtime of programs containing non-linear arithmetic.\paper{ We
  refer to \cite{report} for all proofs.}\report{ All proofs can be found in \cref{app:proofs}.}

\section{Preliminaries}
\label{Preliminaries}

This  section recapitulates preliminaries
for complexity analysis from \cite{brockschmidt2016AnalyzingRuntimeSize,Festschrift}.

\begin{definition}[Atoms and Formulas]\label{Atoms and Formulas}
We fix a set $\VSet$ of variables.  
	The set of \emph{atoms} $\AtomSet(\VSet)$ consists of all inequations $p_1 < p_2$
       for polynomials $p_1,p_2\in\ZZ[\VSet]$.
        $\FormulaSet(\VSet)$ is the set of all propositional \emph{formulas} built from
        atoms $\AtomSet(\VSet)$,  $\land$, and $\lor$. 
\end{definition}
In addition to ``$<$'', we also use ``$\geq$'', ``$=$'', ``$\neq$'', etc.,  and
negations ``$\neg$'' which can be simulated by formulas (e.g., $p_1 \geq p_2$ is
equivalent to $p_2 < p_1 + 1$ for integers).

For integer programs, we use a formalism based on transitions, which also allows us to represent \textbf{while}-programs like \eqref{WhileExample} easily.
Our programs may have \emphit{non-deterministic branching}, i.e., the guards of several
applicable transitions can be satisfied.
Moreover, \emph{non-deterministic sampling} is modeled by \emphit{temporary variables}
whose values are updated arbitrarily in each evaluation step.

\begin{definition}[Integer Program]\label{Integer Program}
$\IntProgram$ is an \emph{integer program}  where
	\begin{itemize}
		\item[$\bullet$] $\PVSet\subseteq\VSet$ is a finite set of \emph{program
                variables},
		     $\VSet \setminus \PVSet$ are \emph{temporary variables} \paper{\vspace{-.3cm}}\report{\vspace{.15cm}}
		\item[$\bullet$] $\LSet$ is a finite set of \emph{locations} with an
                  \emph{initial location} $\location_0\in\LSet$ \vspace{.15cm}
		\item[$\bullet$] $\TSet$ is a finite set of \emph{transitions}.
		      A transition is a 4-tuple $(\location,\guard,\update,\location')$ with a \emph{start location} $\location\in\LSet$,  \emph{target location} $\location'\in\LSet\setminus\braced{\location_0}$,  \emph{guard} $\guard\in\FormulaSet(\VSet)$,
		      and \emph{update function} $\update: \PVSet\rightarrow\ZZ[\VSet]$
                      mapping program variables
                      to update polynomials.
	\end{itemize}
\end{definition}
Transitions $(\location_0,\_,\_,\_)$ are called \emph{initial}.
Note that $\location_0$ has no incoming transitions.

\begin{example}
	\begin{figure}[t]
		\centering
	\hspace*{-.5cm}	\begin{tikzpicture}[->,>=stealth',shorten >=1pt,auto,node distance=3.5cm,semithick,initial text=$ $]
			\node[state,initial] (q0) {$\location_0$};
			\node[state] (q1) [right of=q0,xshift=-2cm]{$\location_1$};
                       			\node[state,draw=none] (h0) [right of=q1, node distance=2cm]{};
			\node[state] (q2) [below of=h0, node distance=1.2cm]{$\location_2$};
			\node[state] (q3) [right of=q1, node distance=4.5cm]{$\location_3$};
			\draw (q0) edge node [text width=1.5cm,align=center] {{\scriptsize
                            $t_0$}} (q1);
			\draw (q1) edge[bend left=75, looseness=.4] node [text
                          width=4cm,align=center,below,yshift=-.1cm] {{\scriptsize $t_1:\guard = x_3 > 0 \wedge x_4 > 0$ \\
			    $\begin{array}{rcl}
                              \update(x_1) &=& x_4 \\
			      \update(x_2) &=& x_5
\end{array}$}} (q3);
			\draw (q3) edge[loop right, min distance=1cm] node [text width=5cm,align=center,yshift=.3cm,xshift=-1.1cm]
                              {{\scriptsize $t_5:\guard = x_1^2 + x_3^5 < x_2 \wedge x_1
                                  \neq 0$ \\ $\begin{array}{rcl}
					\update(x_1) &=& -2\cdot x_1 \\
					\update(x_2) &=& 3\cdot x_2 - 2\cdot x_3^3
\end{array}$}} (q3);
			\draw (q1) edge[bend right] node [text
                          width=4cm,align=center,below, xshift=-1.8cm, yshift=.2cm] {{\scriptsize $t_3:\guard =	-5 \leq x_3 \leq 5\;$}} (q2);
			\draw (q2) edge[bend right] node [text
                          width=4cm,align=center,below, xshift=1.9cm, yshift=.5cm] {{\scriptsize $t_4:\guard = x_4 > 0$ \\
			  $\begin{array}{rcl}
		            \update(x_1) &=& x_4 \\
			    \update(x_2) &=& x_5
\end{array}$}} (q3);
			\draw (q3) edge[bend left=10] node [text width=3cm,align=center]
                              {{\scriptsize $t_2:\update(x_4) = x_4 - 1$}} (q1);
		\end{tikzpicture}
		\caption{An Integer Program with a Nested
                  Self-Loop}\label{fig:twnselfloop}
                \vspace*{-.2cm}
	\end{figure}
Consider the program in \cref{fig:twnselfloop}
with $\PVSet = \{x_i \mid 1 \leq i \leq 5\}$, $\LSet =\linebreak \{\location_i \mid 0 \leq i \leq
3\}$, and $\TSet = \braced{t_i \mid 0 \leq i \leq 5}$, where $t_5$ has non-linear
arithmetic in its guard and update.
We omitted trivial guards, i.e., $\guard = \true$, and identity
        updates of the form $\update(v) = v$. Thus, $t_5$ corresponds to
        the \textbf{while}-program \eqref{WhileExample}.
\end{example}

A \emph{state} is a mapping $\valuation:\VSet\rightarrow\ZZ$,
$\Valuation$ denotes the set of all states, and
$\LSet\times\Valuation$ is the set of \emph{configurations}.
We also apply states to arithmetic expressions $p$ or formulas $\varphi$, where the number $\valuation(p)$ resp.\ the Boolean value $\valuation(\varphi)$ results from replacing each variable $v$ by $\valuation(v)$.
So for a state with $\valuation(x_1) = -8$, $\valuation(x_2) = 55$, and $\valuation(x_3) =
1$,
the expression $x_1^2 + x_3^5$ evaluates to $\valuation(x_1^2 + x_3^5) = 65$ and
the formula $\varphi = (x_1^2 + x_3^5 < x_2)$ evaluates to $\valuation(\varphi) = (65
< 55) = \false$.
From now on, we fix a program $\IntProgram$.

\begin{definition}
	[Evaluation of  Programs]\label{def:Evaluation} For configurations $(\location,\valuation)$, $(\location',\valuation')$ and $t = (\location_t,\guard,\update,\location_{t}')\in\TSet$, $(\location,\valuation)\rightarrow_t(\location',\valuation')$ is an \emph{evaluation}
	step if
 $\location = \location_t$,  $\location' = \location_{t}'$, $\valuation(\guard) =
        \normalfont{\true}$, and $\valuation(\update(v)) = \valuation'(v)$ for all $v\in\PVSet$.
        Let
        $\to_{\TSet} \; =  \, \bigcup_{t \in \TSet} \to_t$, where we also write $\to$ instead of
         $\to_t$ or $\to_{\TSet}$.
Let   $(\location_0,\valuation_0)\rightarrow^k(\location_k,\valuation_k)$
 abbreviate $(\location_0,\valuation_0)\linebreak \rightarrow \ldots
        \rightarrow(\location_k,\valuation_k)$ and let $(\location,\valuation) \rightarrow^*(\location',\valuation')$ if $(\location,\valuation) \rightarrow^k(\location',\valuation')$ for some $k \geq 0$.
\end{definition}

So when denoting states $\valuation$ as tuples $(\valuation(x_1),\ldots,\valuation(x_5))
\in \ZZ^5$
for the program in \cref{fig:twnselfloop},
we have $(\location_0,(5,7,1,1,3)) \to_{t_0}
	(\location_1,(5,7,1,1,3))\to_{t_1}
	(\location_3,(1,3,1,1,3)) \to_{t_5}^3
	(\location_3,(-8,55,1,1,3)) \to_{t_2} \ldots$.
The runtime complexity $\rc(\initial)$ of a program corre\-sponds to the length of the longest evaluation starting in the initial state $\initial$.

\begin{definition}[Runtime Complexity]
 	The \emph{runtime complexity}
  is $\rc\!:\!\Valuation\!\rightarrow\!\NNC$ with $\NNC\!=\!\NN\cup\braced{\omega}$ and $\rc(\initial)\!=\!\sup\braced{k\!\in\!\NN\!\mid\!\exists (\location',\valuation').\, (\location_0,\initial)\!\rightarrow^k\!(\location',\valuation')}$.
\end{definition}

\section{Computing Global Runtime Bounds}
\label{Computing Global Runtime Bounds}

We now introduce our general approach for computing (upper) runtime bounds.
We use weakly monotonically increasing functions as bounds, since
they can easily be ``composed'' (i.e., if $f$ and $g$ increase monotonically, then so does $f \circ g$).

\begin{definition}[Bounds \cite{brockschmidt2016AnalyzingRuntimeSize,Festschrift}]
	\label{def:bounds}
	The set of \emph{bounds}
	$\BoundSet$ is the smallest set with $\overline{\NN} \subseteq \BoundSet$, $\PVSet
        \subseteq \BoundSet$, and $\{b_1+b_2, \, b_1 \cdot b_2, \, k^{b_1}\} \subseteq \BoundSet \text{ for all } k \in \NN$ and $b_1,b_2 \in \BoundSet$.
\end{definition}
A bound constructed from $\NN$, $\PVSet$, $+$, and $\cdot$ is \emph{polynomial}.
So for $\PVSet = \{x,y\}$, we have $\omega$, $x^2$, $x+y$, $2^{x + y} \in \BoundSet$.
Here, $x^2$ and $x+y$ are polynomial bounds.

We measure the size of variables by their absolute values.
For any $\valuation \in \Valuation$,  $\abs{\valuation}$ is\linebreak the state with $\abs{\valuation}(v) = \abs{\valuation(v)}$ for all $v \in \VSet$.
So if $\initial$ denotes the initial state,
then\linebreak $|\initial|$ maps every variable to its initial ``size'', i.e., its initial absolute value.
$\glo: \TSet \rightarrow \BoundSet$ is a \emph{global runtime bound}
if for each transition $t$ and initial state $\initial\in \Valuation$, $\glo(t)$
evaluated in the state $|\initial|$ over-approximates the number of evaluations of $t$ in any run starting in the configuration $(\location_0,\valuation_0)$.
Let $\rightarrow^*_{\TSet} \circ \rightarrow_t$ denote the relation where arbitrary many
evaluation steps are followed by a step with  $t$.

\begin{definition}[Global Runtime Bound  \cite{brockschmidt2016AnalyzingRuntimeSize,Festschrift}]
	\label{def:gloUpperTimeBound}
	The function $\glo: \TSet \rightarrow \BoundSet$ is a \emph{global runtime bound}
	if for all $t \in \TSet$ and all states $\initial \in \Valuation$ we have
	$|\initial|(\glo(t)) \; \geq \; \timeboundterm$.
\end{definition}

For the program in \Cref{fig:twnselfloop}, in \cref{ex:fullExample} we will infer $\glo(t_0) =1$, $\glo(t_i) = x_4$ for $1 \leq i \leq 4$,
and $\glo(t_5) = 8 \cdot x_4 \cdot x_5 + 13006\cdot x_4$.
By adding the bounds for all transitions, a global runtime bound $\glo$ yields an upper bound on the program's runtime complexity.
So for all $\initial\in\Valuation$ we have $|\initial|(\sum_{t\in\TSet}\glo(t)) \geq \rc(\initial).$

For \emph{local runtime bounds}, we consider the \emph{entry transitions} of subsets
$\TSet'\subseteq \TSet$.

\begin{definition}[Entry Transitions \cite{brockschmidt2016AnalyzingRuntimeSize,Festschrift}]
	Let $\emptyset\neq\TSet' \subseteq \TSet$.
Its \emph{entry transi\-tions}  are $ \entry_{\TSet'} = \braced{t \mid t\!=\!(\location,\guard,\update,\location')\!\in\!\TSet\setminus\TSet'\wedge \text{ there is a transition } (\location',\_,\_,\_)\!\in\!\TSet'}$.
\end{definition}
So in \cref{fig:twnselfloop}, we have $\entry_{\TSet \setminus \{ t_0 \}}
	= \{ t_0 \}$ and  $\entry_{\{t_5\}}
	= \{ t_1,t_4 \}$.

In contrast to global runtime bounds, a \emph{local} runtime bound $\loc: \entry_{\TSet'}
\rightarrow \BoundSet$ only takes a subset $\TSet'$ into account.
A \emph{local run} is started by an entry transition $\pret\in\entry_{\TSet'} $ followed by transitions from $\TSet'$.
A \emph{local runtime bound} considers a subset $\TSet'_>\subseteq \TSet'$ and
over-approximates the number of evaluations of any transition from $\TSet'_>$ in an
arbitrary local run of the subprogram with the transitions $\TSet'$.
More precisely, for every $t \in \TSet'_>$, $\loc(r)$ over-approximates the number of
applications of $t$ in any run of $\TSet'$, if $\TSet'$ is entered via  $\pret \in
\entry_{\TSet'}$.
However, local runtime bounds do not consider how often an entry transition from $\entry_{\TSet'}$ is evaluated or how large a variable is when we evaluate an entry transition.
To illustrate that $\loc(\pret)$ is a bound on the number of evaluations of transitions from $\TSet'_>$ after evaluating $\pret$, we often write $\loc(\rightarrow_{\pret}\TSet'_>)$ instead of $\loc(\pret)$.
\begin{definition}[Local Runtime Bound]
	\label{def:locUpperTimeBound}
	Let $\emptyset\neq\TSet'_>\subseteq \TSet'\subseteq\TSet$.
	The function $\loc: \entry_{\TSet'} \rightarrow \BoundSet$ is a \emph{local
        runtime bound} for $\TSet'_>$ w.r.t.\ $\TSet'$ if for all $t \in \TSet'_>$, all
        $\pret\in\entry_{\TSet'}$ with $r = (\location, \_,\_,\_)$, and all
        $\valuation \in \Valuation$ we have
     	$|\valuation|(\loc(\rightarrow_{\pret}\TSet'_>)) \; \geq \;
	 \sup \braced{ k \in \NN \mid \exists\, \initial,
                (\location', \valuation'). \;
(\location_0, \initial) \rightarrow_{\TSet}^* \circ  \rightarrow_{\pret} \, (\location, \valuation) \; (\rightarrow_{\TSet'}^* \circ \rightarrow_t)^k \; (\location', \valuation') }$.
\end{definition}

Our approach is \emph{modular} since it computes local bounds for program
parts
separately. To lift local to global runtime bounds, we use \emph{size bounds} $\Size(t, v)$ to over-approximate the size (i.e., absolute value) of the variable $v$ after evaluating $t$ in any run of the program.
See \cite{brockschmidt2016AnalyzingRuntimeSize} for the automatic computation of size bounds.

\begin{definition}[Size Bound  \cite{brockschmidt2016AnalyzingRuntimeSize,Festschrift}]
	\label{sizebounds}
	The function $\Size: (\TSet \times \PVSet) \rightarrow \BoundSet$ is a \emph{size bound}
	if for all $(t, v) \in \TSet \times \PVSet$ and all states $\initial \in
        \Valuation$ we have $|\initial|(\Size(t, v)) \geq \sup\sizeboundterm$.
\end{definition}

To compute global from local runtime bounds
$\loc(\rightarrow_{\pret}\TSet'_>)$
and size bounds $\Size(\pret,v)$, \cref{thm:time-bound} generalizes the approach of
\cite{brockschmidt2016AnalyzingRuntimeSize,Festschrift}.
Each local run is started by an entry transition $\pret$.
Hence, we use an already computed global runtime bound $\glo(\pret)$
to over-approximate the number of times that such a local run is started.
To over-approximate the size of each variable $v$ when entering the local run,
we instantiate it by the size bound $\Size(\pret,v)$.
So size bounds on previous transitions are needed to compute runtime bounds, and
similarly, runtime bounds are needed to compute size bounds in \cite{brockschmidt2016AnalyzingRuntimeSize}.
For any bound $b$,  ``$b \; [v/ \Size(\pret,v) \mid v \in \PVSet]$'' results from $b$ by replacing every program variable $v$ by $\Size(\pret,v)$.
Here, weak monotonic increase of $b$ ensures that the over-approximation of the variables
$v$ in $b$
by $\Size(\pret,v)$\linebreak indeed also leads to an over-approximation of $b$.
The analysis starts with an \emph{initial} runtime bound $\glo$ and an \emph{initial} size
bound $\Size$ which map all transitions resp.\ all pairs from $\TSet \times \PVSet$ to
$\omega$, except for the transitions $t$ which do not occur in cycles of $\TSet$, where
$\glo(t) = 1$.
Afterwards, $\glo$ and $\Size$ are refined repeatedly, where we alternate between
computing runtime and size bounds.

\begin{theorem}[Computing Global Runtime Bounds]
	\label{thm:time-bound}
	Let $\glo$ be a global runtime bound, $\Size$ be a size bound, and
 $\emptyset\neq \TSet'_> \subseteq \TSet' \subseteq \TSet$ such that $\TSet'$ contains no initial transitions.
	Moreover, let $\loc$ be a local runtime bound for $\TSet'_>$ w.r.t.\ $\TSet'$.
Then  $\glopr$ is also a global runtime bound, where for
        all $t \in \TSet$ we define:
	\[
		\glopr(t)\!=\!		\left\{\begin{array}{ll}
		\!\glo(t),        &\text{if $t\!\in\!\TSet\!\setminus\!\TSet'_>$}\\
	        \!\sum_{\pret \in \entry_{\TSet'}} \glo(\pret)\cdot (\loc(\rightarrow_{\pret}\!\TSet'_>) \left[v/\Size(\pret,v) \mid v\!\in\!\PVSet \right]), &\text{if $t\!\in\!\TSet'_>$}
		\end{array} \right.
	\]
\end{theorem}
\makeproof{thm:time-bound}{
	\begin{myproof}
		We show that for all $t\in\TSet$ and all $\valuation_0 \in \Valuation$ we have
		\begin{align*}
			|\initial|(\glopr(t)) \geq \timeboundterm.\label{eq:sound_timebound}
		\end{align*}
		The case $t \notin \TSet'_>$ is trivial, since then we have $\glopr(t) = \glo(t)$ and $\glo$ is a global runtime bound.
For  $t \in \TSet'_>$, let $(\location_0, \valuation_0) \, (\rightarrow^*_{\TSet} \circ \rightarrow_{t})^k \, (\location', \valuation')$ and we have to show $|\initial|(\glopr(t)) \geq k$.

		If $k = 0$, then we clearly have $|\initial|(\glopr(t)) \geq 0 = k$.
		Hence, we consider $k > 0$.
		We represent the evaluation as follows for numbers $\tilde{k}_i \geq 0$ and $k_i' \geq 1$:
		\begin{IEEEeqnarray*}{llll}
			(\prel_0, \prestate_0) &(\rightarrow^{\tilde{k}_0}_{\TSet \setminus \TSet'} \circ \rightarrow_{\pret_1}) & (\actl_1, \actstate_1) & \rightarrow^{k_1'}_{\TSet'} \\
			(\prel_1, \prestate_1) &(\rightarrow^{\tilde{k}_1}_{\TSet \setminus \TSet'} \circ \rightarrow_{\pret_2}) & (\actl_2, \actstate_2) & \rightarrow^{k_2'}_{\TSet'} \\
			&\hspace*{1.5cm}\vdots & & \\
			(\prel_{m-1}, \prestate_{m-1}) \,& (\rightarrow^{\tilde{k}_{m-1}}_{\TSet \setminus \TSet'} \circ \rightarrow_{\pret_{m}}) \,& (\actl_{m}, \actstate_{m}) \,& \rightarrow^{k_{m}'}_{\TSet'} \\
			(\prel_m, \prestate_m)
		\end{IEEEeqnarray*}
		So for the evaluations from $(\prel_i, \prestate_i)$ to $(\actl_{i+1}, \actstate_{i+1})$ we only use transitions from $\TSet\setminus\TSet'$, and for the evaluations from $(\actl_i, \actstate_i)$ to $(\prel_i, \prestate_i)$ we only use transitions from $\TSet'$.
		Thus, $t$ can only occur in the following finite sequences of evaluation steps:
		\begin{equation}
			\label{SubsetEvaluation}
			(\actl_i, \actstate_i) \rightarrow_{\TSet'} (\actl_{i,1}, \actstate_{i,1}) \rightarrow_{\TSet'} \ldots \rightarrow_{\TSet'} (\actl_{i,k_i'-1}, \actstate_{i,k_i'-1}) \rightarrow_{\TSet'} (\prel_i, \prestate_i).
		\end{equation}
		For every $1 \leq i \leq m$, let $k_i \leq k_i'$ be the number of times that $t$ is used in the evaluation \eqref{SubsetEvaluation}.
		Clearly, we have
		\begin{equation}
			\label{SumDecreasingTransition}
			\sum_{i=1}^{m} k_i = k.
		\end{equation}

		As $\Size$ is a size bound, we have $|\initial| (\Size(\pret_i, v)) \geq |\actstate_i(v)|$ for all $v\in\PVSet$.
		Hence, by the definition of local runtime bounds and as bounds are weakly monotonically increasing functions, we can conclude that
		\begin{equation}
                  \label{gloThmHelp}
			|\initial| (\loc(\rightarrow_{\pret_i} \TSet'_>) \left[v/\Size(\pret_i,v) \mid v \in \PVSet \right]) \; \geq \; |\actstate_i|(\loc(\rightarrow_{\pret_i} \TSet'_>)) \; \geq \; k_i.
		\end{equation}

		Finally, we need to analyze how often such evaluations $(\actl_i, \actstate_i) \rightarrow^*_{\TSet'} (\prel_i, \prestate_i)$ can occur.
		Every entry transition $\pret_i\in\entry_{\TSet'}$ can occur at most $|\initial| (\glo(\pret_i))$ times in the complete evaluation, as $\glo$ is a global runtime bound.
		Thus, we have
		\begin{align*}
			|\initial| (\glopr(t))
			{} = {}
			           & \sum_{\pret \in \entry_{\TSet'}}|\initial|(\glo(\pret))\cdot |\initial|(\loc(\rightarrow_{\pret}\TSet'_>) \left[v/\Size(\pret,v) \mid v \in \PVSet \right]) \\
			{} \geq {} & \sum_{i=1}^m |\initial| (\loc(\rightarrow_{\pret_i} \TSet'_>)\left[v/\Size(\pret_i,v) \mid v \in \PVSet \right] )                                            \\
			{} \geq {} & \sum_{i=1}^m k_i \tag{ by \eqref{gloThmHelp}}                                                                                                                      \\
			{} = {}    & k \tag{ by \eqref{SumDecreasingTransition}}
		\end{align*}
	\end{myproof}
}

\begin{example}
	\label{ex:fullExample}
For the example in \cref{fig:twnselfloop}, we first use
	 $\TSet'_> = \braced{t_2}$ and $\TSet' = \TSet\setminus\braced{t_0}$.
       With the ranking function $x_4$ one obtains
$\loc(\rightarrow_{t_0}\TSet'_>) = x_4$,
since $t_2$ decreases
the value of $x_4$ and no transition increases it.
	Then we can infer the global runtime bound $\glo(t_2) = \glo(t_0)\cdot (x_4\left[v/\Size(t_0,v) \mid v\in\PVSet \right]) = x_4$ as $\glo(t_0) = 1$ (since $t_0$ is evaluated at most once) and $\Size(t_0,x_4) = x_4$ (since $t_0$ does not change any variables).
	Similarly, we can infer  $\glo(t_1) = \glo(t_3) = \glo(t_4) = x_4$.

        For $\TSet'_> = \TSet' = \{t_5\}$, our twn-approach in \Cref{Local Runtime Bounds for TWN-Self-Loops} will infer the
        local runtime bound $\loc: \entry_{\braced{t_5}}
        \rightarrow \BoundSet$ with $\loc(\rightarrow_{t_1}\braced{t_5}) = 4\cdot x_2 + 3$
        and $\loc(\rightarrow_{t_4}\braced{t_5})\linebreak = 4 \cdot x_2 + 4\cdot x_3^3 + 4\cdot
        x_3^5 + 3$  in \Cref{ex:twnloopInLargerProgram}.
	By  \Cref{thm:time-bound} we   obtain the global bound
       \[	\begin{array}{r@{\;\;}c@{\;\;}l}
 \glo(t_5)  &=	 &\glo(t_1)\cdot (\loc(\rightarrow_{t_1}\braced{t_5})[v/\Size(t_1,v) \mid v \in \PVSet	])
  \;  +\\
  &&\glo(t_4)\cdot (\loc(\rightarrow_{t_4}\braced{t_5})[v/\Size(t_4,v) \mid v \in \PVSet])
  \\
                          		           &=& x_4\cdot(4\cdot x_5 + 3) + x_4\cdot
                                                   (4 \cdot x_5 + 4\cdot 5^3 + 4\cdot 5^5
                                                   + 3)\\
                                                   &&
                                                 \hspace*{1.9cm}  \text{(as $\Size(t_1,x_2) = \Size(t_4,x_2) = x_5$ and $\Size(t_4,x_3) = 5$)} \\
		           &=& 8 \cdot x_4 \cdot x_5 + 13006\cdot x_4.
	\end{array}\]
Thus, $\rc(\initial) \in \landau(n^2)$ where $n$ is the largest initial
        absolute value of all program variables.
While the approach of \cite{brockschmidt2016AnalyzingRuntimeSize,Festschrift} was limited to local bounds resulting from ranking
functions, here we need our \cref{thm:time-bound}. It allows us to use
        both local bounds resulting from
        twn-loops (for the non-linear transition $t_5$ where tools based on ranking
        functions cannot infer a bound, see \Cref{Evaluation})
 and
local bounds resulting from ranking functions (for $t_1,\ldots,t_4$, since
our twn-approach of
\cref{Local Runtime Bounds for
  TWN-Self-Loops,Local Runtime Bounds for TWN-Cycles}
is limited to so-called simple cycles and
cannot handle the full program).

In contrast to \cite{brockschmidt2016AnalyzingRuntimeSize,Festschrift}, we allow
different local bounds for different entry transitions in \Cref{def:locUpperTimeBound,thm:time-bound}.
Our example demonstrates that
this can indeed lead to a smaller asymptotic bound
for the whole program:
By distinguishing the cases where $t_5$ is reached via
$t_1$ or $t_4$, we end up with a quadratic bound, because the
local bound
$\loc(\rightarrow_{t_1}\braced{t_5})$ is linear and while $x_3$ occurs with degrees 5 and 3
in
$\loc(\rightarrow_{t_4}\braced{t_5})$, the size bound for $x_3$ is constant after $t_3$
and $t_4$.
  \end{example}

To improve size and runtime bounds repeatedly, we treat the
strongly connected components
(SCCs)\footnote{As usual, a graph is \emph{strongly connected} if there is a path from every node to every other node.
	A \emph{strongly connected component}
	is a maximal strongly connected subgraph.}
of the program  in topological order such that improved bounds for previous transitions
are already available when handling the next SCC.
We first try to infer local runtime bounds by multiphase-linear ranking functions
(see \cite{Festschrift} which also contains a
heuristic for choosing $\TSet'_>$ and $\TSet'$ when using ranking functions).
If ranking functions
do not yield finite local bounds  for all transitions
of the SCC, then we apply the twn-technique
from \cref{Local Runtime Bounds for TWN-Self-Loops,Local Runtime Bounds for TWN-Cycles} on
the remaining unbounded transitions
(see \cref{Local Runtime Bounds for TWN-Cycles} for choosing $\TSet'_>$ and $\TSet'$
in that case).
Afterwards, the global runtime bound is updated according to \cref{thm:time-bound}.

\section{Local Runtime Bounds for Twn-Self-Loops}
\label{Local Runtime Bounds for TWN-Self-Loops}

In \Cref{Termination of Twn-Loops} we recapitulate twn-loops and their termination in our setting. Then\linebreak in
\Cref{Runtime Bounds for Twn-Loops} we present a (complete) algorithm to infer polynomial
runtime bounds\linebreak for all terminating twn-loops.
Compared to
\cite{hark2020PolynomialLoopsTermination}, we increased its precision consider\-ably by
computing bounds that take the different roles of the variables into account\linebreak
and by using over-approximations to
remove monomials.
Moreover, we show how our algorithm can be used to infer local runtime bounds for twn-loops occurring in integer programs.
\cref{Local Runtime Bounds for TWN-Cycles}  will show that our algorithm can also be
applied to infer runtime bounds for larger cycles in programs instead of just self-loops.

%

\subsection{Termination of Twn-Loops}\label{Termination of Twn-Loops}

\Cref{def:twn-loop} extends the definition of twn-loops in
\cite{hark2020PolynomialLoopsTermination,frohn2020TerminationPolynomialLoops} by
an initial transition and an update-invariant.
Here,  $\invariant$ is an \emph{update-invariant} if $\models \invariant\rightarrow
        \update(\invariant)$ where $\update$ is the update of the transition (i.e.,  invariance
        must hold independent of the guard).

\begin{definition}[Twn-Loop]\label{def:twn-loop}
	An integer program $(\PVSet, \LSet, \location_0, \TSet)$ is a \emph{triangular weakly non-linear loop (twn-loop)}
	if $\PVSet = \{ x_1,\ldots,x_\indv\}$ for some $\indv \geq 1$, $\LSet = \{
        \location_0, \location \}$, and $\TSet = \{ t_0, t \}$ with $t_0 = (\location_0,
        \psi, \mathrm{id}, \location)$ and $t = (\location, \varphi, \update, \location)$
        for some $\psi, \varphi\linebreak \in \FormulaSet(\PVSet)$
with  $\models \invariant \rightarrow \update(\invariant)$, where $\mathrm{id}(v) = v$ for
all $v \in \PVSet$, and for all $1 \leq i \leq d$\linebreak we have $\update(x_i) = c_i
\cdot x_i + p_i$ for some $c_i \in \ZZ$ and some polynomial $p_i \in
\ZZ[x_{i+1},\linebreak \ldots,x_\indv]$.
	We often denote the loop by
        $\IntLoop$ and refer to $\invariant$, $\guard$, $\update$ as its
        \mbox{(update-)}\linebreak invariant, guard, and update, respectively.
	If $c_i \geq 0$ holds for all $1 \leq i \leq d$, then the program is a
        \emph{non-negative} triangular weakly non-linear loop
        \emph{(tnn-loop)}.
\end{definition}

\begin{example}
	\label{ex:twn}
        The program consisting of the initial transition $(\location_0, \true,
        \mathrm{id}, \location_3)$ and the self-loop $t_5$ in \cref{fig:twnselfloop}
        is a twn-loop (corresponding to the \textbf{while}-loop \eqref{WhileExample}).
	This loop terminates as every iteration increases $x_1^2$ by a factor of $4$ whereas
       $x_2$ is only tripled. Thus,
 $x_1^2 + x_3^5$ eventually outgrows the value of $x_2$.
\end{example}

To transform programs into twn- or tnn-form, one can combine subsequent transitions by \emph{chaining}.
Here, similar to states $\sigma$, we also apply the update $\update$ to polynomials and formulas by replacing each program variable $v$ by $\update(v)$.

\begin{definition}[Chaining]
  Let $t_1,\ldots,t_n$ be a sequence of transitions without temporary variables where
 $t_i = (\location_i, \guard_i, \update_i, \location_{i+1})$
  for all $1 \leq i \leq
        n-1$, i.e., the
      target location of $t_i$ is the start location of $t_{i+1}$.
We may have $t_i = t_j$ for $i \neq j$, i.e., a transition may occur several times in the sequence.
Then the transition $t_1 \chain \ldots \chain t_n = (\location_1, \guard, \update,
\location_{n+1})$  results from
\emph{chaining} $t_1,\ldots,t_n$  where
	\[
		\begin{array}{rcl}
			\guard     & = & \guard_1 \, \land \, \update_1(\guard_2) \, \land
                        \, \update_1(\update_2(\guard_3))  \, \land \, \ldots \,   \land
                        \, \update_1(\ldots\update_{n-1}(\guard_n)\ldots) \\
			\update(v) & = & \update_1(\ldots\update_n(v)\ldots) \text{ for all $v \in \PVSet$, i.e., $\update = \update_1 \circ \ldots \circ \update_n$.}
		\end{array}
	\]
\end{definition}

Similar to \cite{frohn2020TerminationPolynomialLoops,hark2020PolynomialLoopsTermination},
we can restrict ourselves to tnn-loops, since
chaining transforms any twn-loop $L$ into a tnn-loop
$L \chain L$.
Chaining preserves the termination behavior, and a bound on  $L
\chain L$'s runtime can be transformed into a bound for $L$.

\begin{lemma}[Chaining Preserves Asymptotic Runtime, see \protect{\cite[Lemma 18]{hark2020PolynomialLoopsTermination}}]
	\label{Chaining Preserves Asymptotic Runtime}
        For the twn-loop $L= \IntLoop$ with the
        transitions  $t_0 = (\location_0, \psi, \mathrm{id}, \location)$,
        $t = (\location, \varphi, \update, \location)$, and runtime complexity $\rc_L$,
        the program $L \chain L$
	 with the transitions $t_0$ and $t \chain t = (\invariant,
        \guard\wedge \update(\guard), \update\circ\update)$ is a tnn-loop. For its
        runtime complexity $\rc_{L \chain L}$, we have $
2 \cdot
        \rc_{L \chain L}(\valuation) \;
        \leq \;
        \rc_{L}(\valuation) \; \leq  \; 2 \cdot \rc_{L \chain L}(\valuation) + 1$ for all $\valuation\in\Valuation$.
\end{lemma}

\begin{example}\label{ex:tnn}
  The program of \Cref{ex:twn} is only a twn-loop and not a tnn-loop as $x_1$ occurs with a negative coefficient $-2$ in its own update.
  Hence,  we chain the loop and consider $t_5\chain t_5$.
	The update of  $t_5\chain t_5$ is $(\update \circ
        \update)(x_1) = 4\cdot x_1$,  $(\update \circ
        \update)(x_2) = 9\cdot x_2 - 8\cdot x_3^3$, and
         $(\update \circ
        \update)(x_3) = x_3$.
        To ease the presentation, in this example we will keep the guard $\guard$
        instead of using $\guard\wedge \update(\guard)$
	(ignoring $\update(\guard)$ in the conjunction of the guard does not decrease the runtime
        complexity).
 \end{example}

Our algorithm starts with computing a closed form for the loop update,
which
describes the values of the program variables after $n$ iterations of the loop.
Formally, a tuple of arithmetic expressions $\cl{\vec{x}} = (\cl{x_1}, \ldots, \cl{x_\indv})$ over the variables $\vec{x} = (x_1,\ldots,x_\indv)$ and the distinguished variable $n$ is a \emph{(normalized) closed form} for the update $\update$ with \emph{start value $n_0 \geq 0$} if for all $1 \leq i \leq \indv$ and all $\sigma: \{x_1,\ldots,x_\indv,n\} \to \ZZ$ with $\sigma(n) \geq n_0$, we have $\sigma(\cl{x_i}) = \sigma(\update^n(x_i))$.
As shown in \cite{frohn2020TerminationPolynomialLoops,hark2020PolynomialLoopsTermination,frohn2019TerminationTriangularInteger}, for tnn-loops such a normalized closed form and the start value $n_0$ can be computed
by handling one variable after the other, and these normalized closed forms can be represented as so-called \emph{normalized poly-exponential expressions}.
Here, $\NN_{\geq m}$ stands for $\{ x \in \NN \mid x \geq m \}$.

\begin{definition}[Normalized Poly-Exponential Expression  \cite{frohn2019TerminationTriangularInteger,frohn2020TerminationPolynomialLoops,hark2020PolynomialLoopsTermination}]
	Let $\PVSet = \{ x_1,\ldots,x_\indv \}$.
	Then we define the set of all \emph{normalized poly-exponential expressions}
	by
	$\NNPPEE = \{ \sum_{j=1}^\ell p_j \cdot n^{a_j} \cdot b_j^n \mathrel{\Big|}
        \ell, a_j\in \NN, \; p_j\in\QQ[\PVSet], \; b_j\in\NN_{\geq 1} \}$.
\end{definition}

\begin{example}
	\label{ex:closed form}
	A normalized closed form (with start value $n_0 = 0$) for the tnn-loop in \cref{ex:tnn} is
        $\cl{x_1} = x_1 \cdot 4^n$, $\cl{x_2} = (x_2 - x_3^3) \cdot 9^n  + x_3^3$, and
        $\cl{x_3} = x_3$.
\end{example}

Using the normalized closed form,
similar to \cite{frohn2020TerminationPolynomialLoops} one can represent non-termina\-tion of a tnn-loop $\IntLoop$ by the formula
\begin{equation}
	\label{fml ev non-term}
	  \exists\, \xvec\in \ZZ^d, \;  m\in \NN. \; \forall n \in \NN_{\geq
            m}. \; \invariant \land \guard[\vec{x}/\clExp{\vec{x}}{n}].
\end{equation}
Here, $\guard[\vec{x}/\clExp{\vec{x}}{n}]$ means that each variable $x_i$ in $\guard$ is replaced by $\clExp{x_i}{n}$.
Since $\invariant$ is an update-invariant, \pagebreak[2] if $\invariant$ holds, then
$\invariant[\vec{x}/\clExp{\vec{x}}{n}]$ holds as well for all
 $n \geq n_0$. Hence, whenever
$\forall n \in \NN_{\geq
            m}. \; \invariant \land \guard[\vec{x}/\clExp{\vec{x}}{n}]$
holds, then  $\clExp{\vec{x}}{\max\{n_0,m\}}$ witnesses non-termination.
Thus, invalidity  of \eqref{fml ev non-term} is equivalent to termination of the loop.

Normalized poly-exponential expressions have the advantage that it is always clear which addend determines their asymptotic growth when increasing $n$.
So as in \cite{frohn2020TerminationPolynomialLoops}, \eqref{fml ev non-term} can be
transformed into an existential formula and we use an SMT solver\linebreak to prove its
invalidity in order to prove termination of the loop.
As shown in  \cite[Thm.\ 42]{frohn2020TerminationPolynomialLoops},
non-termination of twn-loops over $\ZZ$ is semi-decidable and deciding termination is \textsf{Co-NP}-complete if the loop is linear
and the eigenvalues of the update matrix are rational.

\subsection{Runtime Bounds for Twn-Loops via Stabilization Thresholds}\label{Runtime Bounds for Twn-Loops}

As observed in \cite{hark2020PolynomialLoopsTermination}, since the closed forms for
tnn-loops are poly-exponential expressions that are weakly monotonic in $n$, every
tnn-loop $\IntLoop$ \emph{stabilizes} for each \pagebreak[2] input $\vec{e} \in \ZZ^d$.
So there is a number of loop iterations (a \emph{stabilization
threshold} $\sth_{\IntLoop}(\vec{e})$), such that the truth value of the loop guard
$\guard$ does not change anymore when performing further loop iterations.
Hence, the runtime of every terminating tnn-loop is bounded by its stabilization threshold.

\begin{definition}[Stabilization Threshold]
	Let $\IntLoop$ be a tnn-loop with $\PVSet\linebreak = \{ x_1,\ldots, x_\indv\}$.
	For each $\vec{e} = (e_1,\ldots,e_\indv) \in \ZZ^d$, let $\valuation_{\vec{e}} \in\Valuation$  with $\valuation_{\vec{e}}(x_i) = e_i$ for all $1 \leq i \leq \indv$.
Let $\InvSet \subseteq \ZZ^d$ such that $\vec{e} \in \InvSet$
        iff
$\valuation_{\vec{e}}(\invariant)$  holds.
        Then $\sth_{\IntLoop}: \ZZ^d \to \NN$ is the \emph{stabilization threshold} of  $\IntLoop$ if for all $\vec{e} \in \InvSet$,
  $\sth_{\IntLoop}(\vec{e})$ is the smallest number such that  $\valuation_{\vec{e}}\left( \;
        \update^{n}(\guard) \leftrightarrow \update^{\sth_{\IntLoop}(\vec{e})}(\guard)\;
        \right)$ holds for
 all $n \geq
        \sth_{\IntLoop}(\vec{e})$.
\end{definition}
For the tnn-loop from \Cref{ex:tnn}, it will turn out that
$2 \cdot x_2 + 2\cdot x_3^3 + 2\cdot x_3^5 + 1$
is an upper bound on its stabilization threshold, see \Cref{exa:twnselfloops_global}.

To compute such upper bounds on a tnn-loop's stabilization threshold (i.e., upper bounds
on its runtime if the loop is terminating), we now present a
construc\-tion based on \emph{monotonicity thresholds}, which are computable \cite[Lemma 12]{hark2020PolynomialLoopsTermination}.

\begin{definition}[Monotonicity Threshold \cite{hark2020PolynomialLoopsTermination}]
	Let $(\ub,\ud), (\lb,\ld)\in\NN^2$ such that $(\ub,\ud)\lex(\lb,\ld)$ (i.e., $\ub > \lb$ or both $\ub = \lb$ and $\ud > \ld$).
	For any $k\in\NN_{\geq 1}$, the $k$-\emph{monotonicity threshold} of $(\ub,\ud)$
        and $(\lb,\ld)$ is the smallest $n_0\in\NN$ such that for all $n\geq n_0$ we have $n^{\ud}\cdot \ub^n > k\cdot n^{\ld} \cdot \lb^n$.
\end{definition}
For example, the $1$-monotonicity threshold of $(4,0)$ and $(3,1)$ is $7$ as
the largest root of $f(n) = 4^n - n \cdot 3^n$ is approximately $6.5139$.

Our procedure again instantiates the variables of the loop guard $\guard$ by the
normalized closed form $\clExp{\vec{x}}{n}$ of the loop's update.
However, in the poly-exponential expressions $\sum_{j=1}^\ell p_j \cdot n^{a_j} \cdot
b_j^n$ resulting from $\guard[\vec{x}/\clExp{\vec{x}}{n}]$, the
corresponding technique of \cite[Lemma
  21]{hark2020PolynomialLoopsTermination}
over-approximated the polynomials $p_j$ by a polynomial that did not distinguish the effects of the different  variables
$x_1,\ldots,x_{\indv}$. Such an
over-approximation is only
        useful for a direct asymptotic bound on the runtime of the twn-loop, but it
     is too coarse for a useful \emph{local}
        runtime bound within the complexity analysis of a larger program.
For instance,  in \Cref{ex:fullExample} it is crucial to obtain
local bounds like $4 \cdot x_2 + 4\cdot x_3^3 + 4\cdot
        x_3^5 + 3$ which indicate that only the variable $x_3$ may influence the runtime with an
exponent of $3$ or $5$.
Thus, if the size of $x_3$ is bound by a constant, then the resulting global bound
\pagebreak[2] becomes linear.

So we now improve precision
and over-approximate the polynomials $p_j$ by the\linebreak polynomial $\sqcup \{ p_1, \ldots, p_{\ell}\}$ which contains every monomial $x_1^{e_1}  \cdot \ldots \cdot x_\indv^{e_\indv}$ of $\{
	p_1,\ldots,\linebreak p_{\ell}\}$, using the absolute value of the largest coefficient with which the monomial occurs in $\{
	p_1, \ldots, p_{\ell}\}$.
        Thus, $\sqcup \{ x_3^3-x_3^5,  x_2 -x_3^3\} = x_2 + x_3^3 + x_3^5$.
        In the following let $\vec{x} = (x_1,\ldots,x_\indv)$,
        and for $\vec{e} = (e_1,\ldots,e_d) \in \NN^d$,  $\vec{x}^{\vec{e}}$ denotes
 $x_1^{e_1} \cdot  \ldots \cdot x_\indv^{e_\indv}$.

\begin{definition}[Over-Approximation of Polynomials]
	\label{Over-Approximation of Polynomials}
Let $p_1, \ldots, p_\ell \in \ZZ[\vec{x}]$, and for all $1 \leq j \leq \ell$, let $\mathcal{I}_j \subseteq (\ZZ \setminus \{0\}) \times \NN^\indv$ be the \emph{index set}
	of the polynomial\linebreak $p_j$ where $p_j = \sum_{(c,\vec{e})\in \mathcal{I}_j}
	c \cdot\vec{x}^{\vec{e}}$ and there are no $c \neq c'$ with $(c,\vec{e}),
        (c',\vec{e})\in \mathcal{I}_j$.  For all\linebreak
        $\vec{e}  \in \NN^d$  we define $c_{\vec{e}} \in \NN$ with $c_{\vec{e}}
        = \max \{ |c| \mid (c,\vec{e}) \in \mathcal{I}_1 \cup \ldots \cup
        \mathcal{I}_\ell \}$, where $\max \emptyset = 0$. Then  the
        \emph{over-approximation} of  $p_1, \ldots, p_\ell$ is $\sqcup \{
        p_1, \ldots, p_{\ell}\} = \sum_{\vec{e}  \in \NN^d}
	c_{\vec{e}} \cdot \vec{x}^{\vec{e}}$.
        \end{definition}
Clearly, $\sqcup \{ p_1, \ldots, p_{\ell}\}$ indeed over-approximates the absolute value of each $p_j$.

\begin{corollary}[Soundness of $\sqcup \{ p_1, \ldots, p_{\ell}\}$]
	\label{Soundness of Over-approximation}
	For all $\sigma: \{ x_1,\ldots,x_\indv\} \to \ZZ$ and all $1 \leq j \leq \ell$, we have $|\sigma|(\sqcup \{ p_1, \ldots, p_{\ell}\}) \geq |\sigma(p_j)|$.
\end{corollary}

A drawback  is that $\sqcup \{ p_1, \ldots, p_{\ell}\}$
considers all monomials and
to obtain
weakly\linebreak monotonically increasing bounds from $\BoundSet$,
it uses the
absolute values of their coef\-ficients. This can lead to
polynomials of unnecessarily
high degree. To improve the precision of the resulting bounds, we now allow to over-approximate the poly-exponential
expressions  $\sum_{j=1}^\ell p_j \cdot n^{a_j} \cdot b_j^n$ which result from instantiating the variables of the loop guard by the
closed form. For this over-approximation, we take
the invariant
$\invariant$ of the tnn-loop into account.
So while \eqref{fml ev non-term} showed that update-invariants $\invariant$
can restrict the sets of possible witnesses for non-termination and thus simplify the termination proofs of
twn-loops, we now show that preconditions $\invariant$
can also be useful to improve the bounds on
twn-loops.

More precisely, \Cref{def:non-monotonic kernel} allows us to replace addends  $p \cdot
n^a\cdot b^n$ by   $p \cdot
n^{i}\cdot {j}^n$ where $(j,i) \lex (b,a)$ if the monomial $p$ is always positive (when the
precondition $\invariant$ is fulfilled)
and where  $(b,a) \lex (i,j)$ if $p$ is always non-positive.

\begin{definition}[Over-Approximation of Poly-Exponential Expressions]\label{def:non-monotonic kernel}Let $\invariant
        \in \FormulaSet(\PVSet)$ and
	let $npe = \sum_{(p,a,b)\in\Lambda} \; p \cdot n^a\cdot b^n\in\NNPPEE$
        where $\Lambda$ is a set of tuples $(p,a,b)$ containing a monomial\footnote{Here,
        we consider monomials of the form $p = c \cdot x_1^{e_1} \cdot \ldots \cdot
        x_d^{e_d}$ with coefficients $c \in \QQ$.}
        $p$
        and two numbers $a,b\in\NN$. Here, we may have $(p,a,b), (p',
        a, b) \in \Lambda$ for $p \neq p'$.
	Let $\Delta,\Gamma \subseteq \Lambda$
        such that $\models\psi\rightarrow (p > 0)$   holds for all
        $(p,a,b)\in\Delta$
and $\models\psi\rightarrow (p \leq 0)$
        holds for all
        $(p,a,b)\in\Gamma$.\footnote{$\Delta$ and $\Gamma$ do not have to
        contain
\emph{all} such tuples, but can be (possibly empty) subsets.}
        Then
$$\mbox{$\kernel{npe}{\invariant}{\Delta,\Gamma} = \sum\nolimits_{(p,a,b)\in\Delta\uplus\Gamma}
\; p\cdot n^{i_{(p,a,b)}} \cdot j^n_{(p,a,b)} +
\sum\nolimits_{(p,a,b)\in\Lambda\setminus(\Delta\uplus\Gamma)} \; p \cdot n^a\cdot b^n$}$$
is an \emph{over-approximation} of $npe$ if
       $i_{(p,a,b)},
j_{(p,a,b)} \in \NN$
are numbers such that
 $(j_{(p,a,b)}, i_{(p,a,b)}) \lex (b,a)$ holds if
$(p,a,b)\in \Delta$ and  $(b,a) \lex (j_{(p,a,b)}, i_{(p,a,b)})$ holds if
$(p,a,b)\in \Gamma$. Note that $i_{(p,a,b)}$ or $j_{(p,a,b)}$ can also be $0$.
\end{definition}

\begin{example}\label{ex:kernel}
 Let $npe = q_3 \cdot 16^n + q_2 \cdot 9^n + q_1
= q_3 \cdot 16^n + q_2' \cdot 9^n+ q_2'' \cdot 9^n + q_1' + q_1''$,
where $q_3 = -x_1^2$,
$q_2 = q_2' + q_2''$,
$q_2' = x_2$,
$q_2'' =
- x_3^3$,
$q_1 = q_1' + q_1''$,
$q_1' = x_3^3$,
$q_1'' =
-x_3^5$,
and
$\invariant = (x_3 > 0)$. We can choose  $\Delta = \{ (x_3^3, 0, 1) \}$
        since $\models \invariant \to (x_3^3 > 0)$ and
        $\Gamma =  \{ (-x_3^5, 0, 1) \}$ since $\models \invariant \to (-x_3^5 \leq
        0)$. Moreover, we choose $j_{(x_3^3, 0, 1)} = 9$, $i_{(x_3^3, 0, 1)} = 0$, which is
        possible since $(9,0) \lex (1,0)$. Similarly, we choose $j_{(-x_3^5, 0, 1)} = 0$,
        $i_{(-x_3^5, 0, 1)} = 0$, since $(1,0) \lex (0,0)$. Thus, we replace  $x_3^3$ and  $-x_3^5$
        by the larger addends $x_3^3 \cdot 9^n$ and $0$. The motivation for the latter is
        that this removes all addends with exponent 5 from $npe$. The motivation for
        the former is that then, we have both the addends
        $-x_3^3 \cdot 9^n$ and  $x_3^3 \cdot 9^n$ in the expression which cancel out,
        i.e., this removes all addends with exponent 3. Hence, we obtain
 $\kernel{npe}{\invariant}{\Delta,\Gamma} = p_2 \cdot 16^n + p_1 \cdot 9^n$
        with $p_2 = -x_1^2$ and $p_1 = x_2$. To find a suitable
        over-approximation which removes addends with high exponents,
        our implementation uses a
        heuristic for the choice of $\Delta$, $\Gamma$, $i_{(p,a,b)}$, and
        $j_{(p,a,b)}$.
\end{example}

The following lemma shows the soundness of the over-approximation $\kernel{npe}{\invariant}{\Delta,\Gamma}$.

\begin{lemma}[Soundness of $\kernel{npe}{\invariant}{\Delta,\Gamma}$]
	\label{lem:non-monotonic kernel}
	Let $\invariant$, $npe$, $\Delta$, $\Gamma$, $i_{(p,a,b)}$,
        $j_{(p,a,b)}$, and $\kernel{npe}{\invariant}{\Delta,\Gamma}$
        be as in  \cref{def:non-monotonic kernel}, and let $D_{\kernel{npe}{\invariant}{\Delta,\Gamma}} =$
     \[\begin{array}{rl}    \max(&
        \braced{1\text{-monotonicity threshold of } (j_{(p,a,b)}, i_{(p,a,b)})
          \text{ and } (b,a) \mid (p,a,b)\in\Delta}
        \cup\\
         &\braced{1\text{-monotonicity threshold of } (b,a)
          \text{ and }  (j_{(p,a,b)}, i_{(p,a,b)}) \mid
          (p,a,b)\in\Gamma }\;\;).
        \end{array}\]
     Then for all $\vec{e} \in \InvSet$
     and all $n \geq D_{\kernel{npe}{\invariant}{\Delta,\Gamma}}$, we have
 $\valuation_{\vec{e}}(\kernel{npe}{\invariant}{\Delta,\Gamma}) \geq
  \valuation_{\vec{e}}(npe)$.
\end{lemma}
\makeproof{lem:non-monotonic kernel}{
  \begin{myproof}
For
 all $n \geq D_{\kernel{npe}{\invariant}{\Delta,\Gamma}}$ we have
\[ \begin{array}{rcll}
  \valuation_{\vec{e}}(\, \kernel{npe}{\invariant}{\Delta,\Gamma} \,) &=&
   \multicolumn{2}{l}{\valuation_{\vec{e}}(\,\sum_{(p,a,b)\in\Delta\uplus\Gamma}
 \;    p\cdot n^{i_{(p,a,b)}} \cdot j^n_{(p,a,b)} +}\\
      &&
     \multicolumn{2}{l}{\phantom{\valuation_{\vec{e}}(}\sum_{(p,a,b)\in\Lambda\setminus(\Delta\uplus\Gamma)} \;
       p \cdot n^a\cdot b^n \,)}\\
&=&  \valuation_{\vec{e}}(\,npe &+ \sum_{(p,a,b)\in\Delta} \; p \cdot
   (n^{i_{(p,a,b)}} \cdot j^n_{(p,a,b)} - n^a \cdot b^n)\\
   &&&-
\sum_{(p,a,b)\in\Gamma} \; p \cdot
(n^a \cdot b^n - n^{i_{(p,a,b)}} \cdot j^n_{(p,a,b)})\,)\\
&\geq&  \valuation_{\vec{e}}(\,npe &+ \sum_{(p,a,b)\in\Delta} \; p  -
\sum_{(p,a,b)\in\Gamma} \; p \,)\\
&\geq&   \multicolumn{2}{l}{\valuation_{\vec{e}}(\, npe \,).}
\end{array}\]
	\end{myproof}
}

For any terminating tnn-loop $(\invariant, \guard, \update)$, \cref{thm:complexity}
now uses the new concepts of\linebreak \cref{Over-Approximation of Polynomials,def:non-monotonic kernel}
to compute a polynomial $\sthb$ which is an upper bound on the loop's stabilization threshold (and hence, on its runtime).
      For any atom $\alpha = (s_1 < s_2)$ (resp.\ $s_2-s_1 > 0$) in the loop
      guard $\guard$, let $npe_\alpha\in\NNPPEE$ be
      a poly-exponential expression  which
results from multiplying $(s_2-s_1)[\vec{x}/{\mbox{\normalfont{$\cl{\vec{x}}$}}}]$ with the least common multiple of all denominators occurring in $(s_2-s_1)[\vec{x}/{\mbox{\normalfont{$\cl{\vec{x}}$}}}]$. Since the loop is
        terminating, for some of these atoms
        this expression will become non-positive for large enough $n$ and our
        goal is to compute bounds on their corresponding stabilization thresholds. First,
        one can replace $npe_\alpha$ by an
        over-approximation $\kernel{npe_\alpha}{\invariant'}{\Delta,\Gamma}$
        where $\invariant' = (\invariant \land \guard)$ consid\-ers both the
        invariant $\invariant$  and the
        guard $\guard$.
Let $\InvSet' \subseteq \ZZ^d$ such that $\vec{e} \in \InvSet'$
        iff $\valuation_{\vec{e}}(\invariant')$ holds.
        By \Cref{lem:non-monotonic kernel} (i.e.,  $\valuation_{\vec{e}}(\kernel{npe_\alpha}{\invariant'}{\Delta,\Gamma}) \geq
        \valuation_{\vec{e}}(npe_\alpha)$  for all
$\vec{e} \in \InvSet'$), it suffices to compute a bound on the
 stabilization threshold of   $\kernel{npe_\alpha}{\invariant'}{\Delta,\Gamma}$ if it
 is always non-positive for large enough $n$, because if
 $\kernel{npe_\alpha}{\invariant'}{\Delta,\Gamma}$ is non-positive, then so is
 $npe_\alpha$.
We say that an over-approximation   $\kernel{npe_\alpha}{\invariant'}{\Delta,\Gamma}$ is
\emph{eventually non-positive} iff whenever
$\kernel{npe_\alpha}{\invariant'}{\Delta,\Gamma} \neq npe_\alpha$, then one can show
 that    for all $\vec{e} \in \InvSet'$,
         $\valuation_{\vec{e}}(\kernel{npe_\alpha}{\invariant'}{\Delta,\Gamma})$ is
         always non-positive for large enough $n$.\footnote{\label{eventual non-positivity}This can be shown similar to the proof of \eqref{fml ev non-term}
for (non-)termination of the loop. Thus, we transform $\exists\, \xvec\in \ZZ^d, \;  m\in \NN. \; \forall n \in \NN_{\geq
  m}. \; \invariant' \land  \kernel{npe_\alpha}{\invariant'}{\Delta,\Gamma} > 0$
 into an existential formula as in \cite{frohn2020TerminationPolynomialLoops}
and try to prove its invalidity by an SMT solver.}
         Using over-approximations
 $\kernel{npe_\alpha}{\invariant'}{\Delta,\Gamma}$
      can be advantageous because
 $\kernel{npe_\alpha}{\invariant'}{\Delta,\Gamma}$ may contain less monomials than
 $npe_\alpha$ and thus, the construction $\sqcup$ from \Cref{Over-Approximation of Polynomials} can yield a polynomial of
 lower degree.
So although  $npe_\alpha$'s stabilization
 threshold might be smaller than the one of
 $\kernel{npe_\alpha}{\invariant'}{\Delta,\Gamma}$, our technique might compute a
 smaller bound on the stabilization threshold when considering
$\kernel{npe_\alpha}{\invariant'}{\Delta,\Gamma}$ instead of
$npe$.

\begin{theorem}[Bound on Stabilization Threshold]
	\label{thm:complexity}
	Let $L=(\invariant, \guard, \update)$ be a terminating tnn-loop, let
$\invariant' = (\invariant \land \guard)$,
        and let
        {{\normalfont{$\cl{\vec{x}}$}}} be a normalized closed form for $\update$ with start value $n_0$.
	For every atom $\alpha = (s_1 < s_2)$ in $\guard$,
        let $\kernel{npe_\alpha}{\invariant'}{\Delta,\Gamma}$ be an eventually non-positive
        over-approximation
        of $npe_\alpha$ and let $D_\alpha =
        D_{\kernel{npe_\alpha}{\invariant'}{\Delta,\Gamma}}$.

	If $\kernel{npe_\alpha}{\invariant'}{\Delta,\Gamma} = \sum_{j=1}^{\ell} p_j \cdot
        n^{a_j} \cdot b_j^n$ with $p_j \neq 0$ for all $1 \leq j \leq \ell$ and
        $(b_\ell,a_\ell) \lex \ldots\lex (b_1,a_1)$, then let $C_\alpha = \max\{1, N_2, M_2,
        \ldots, N_\ell, M_\ell \}$, where we have:
\[  \mbox{\small $M_j\!=\!\left\{
\begin{array}{ll}
	0,    & \text{if $b_j = b_{j-1}$} \\
        \text{1-monotonicity threshold of} &\\
        \text{$\;\;\;(b_j,a_j)$  and  $(b_{j-1}, a_{j-1} + 1)$}, & \text{if $b_j > b_{j-1}$}
\end{array}
\right. \quad\;\;
N_j\!=\!\left\{
\begin{array}{ll}
			1,              & \text{if $j = 2$} \\
			mt',            & \text{if $j = 3$} \\
			\max\{mt,mt'\}, & \text{if $j > 3$}
		\end{array}
\right.$}
\]
Here, $mt'$ is the $(j-2)$-monotonicity threshold of $(b_{j-1},a_{j-1})$ and
$(b_{j-2},a_{j-2})$ and
$mt= \max\{1\text{-monotonicity threshold of } (b_{j-2},a_{j-2}) \text{
  and } (b_i,a_i) \mid 1 \leq i \leq j-3\}$.\linebreak
	        Let $Pol_\alpha = \braced{p_1,\ldots, p_{\ell-1}}$,
$Pol = \bigcup_{\text{atom $\alpha$ occurs in $\guard$}} Pol_\alpha$, $C = \max \{C_\alpha
                \mid$ atom $\alpha$ occurs in $\guard \}$,  $D = \max \{D_\alpha \mid \text{atom $\alpha$ occurs in $\guard$}\}$, and $\sthb \in \ZZ[\vec{x}]$ with $\sthb = 2 \cdot \sqcup Pol + \max\{ n_0, C, D\}$.
	Then
      for all  $\vec{e} \in \InvSet'$, we have	$|\valuation_{\vec{e}}|(\sthb)  \geq \sth_{(\invariant,\guard,\update)}(\vec{e})$.
	If the tnn-loop has the initial transition $t_0$ and looping transition $t$, then
        $\glo(t_0) = 1$ and $\glo(t) = \sthb$ is a global runtime
        bound for $L$.
\end{theorem}
\makeproof{thm:complexity}{
	\begin{myproof}
	Let
        $\vec{e} \in \InvSet'$.
	We first prove that
        \begin{equation}
          \label{complexityClaim}
          \max\{ n_0, C_\alpha, D_\alpha, 2 \cdot |\valuation_{\vec{e}}|(
          \sqcup Pol_\alpha) \} \; \geq \; \sth_{(\invariant,a,\update)}(\vec{e})
          \end{equation}
holds for all those atoms $\alpha$ occurring in $\guard$ where
$\valuation_{\vec{e}}(\update^n(\alpha)) = \false$ for all large enough $n$. To this end,
                 we show that  $\valuation_{\vec{e}}(\update^{n}(\alpha)) =\false$ for all $n \geq \max\{ n_0, C_\alpha, D_\alpha,\linebreak 2 \cdot |\valuation_{\vec{e}}|( \sqcup Pol_\alpha) \}$.
		Note that for $n \geq n_0$, by the definition of normalized closed forms,
                we have $\valuation_{\vec{e}}(npe_\alpha) \leq 0$ iff
                $\valuation_{\vec{e}}(\update^{n}(\alpha)) = \false$.
                Thus, we know that  $\valuation_{\vec{e}}(npe_\alpha) \leq 0$ for all large
                enough $n$ and we want to show that it holds for all
                $n \geq \max\{ C_\alpha, D_\alpha, 2 \cdot |\valuation_{\vec{e}}|( \sqcup Pol_\alpha) \}$.

		For all $n \geq D_\alpha$ we have $\valuation_{\vec{e}}(npe_\alpha) \leq \valuation_{\vec{e}}(\kernel{npe_\alpha}{\invariant'}{\Delta,\Gamma})$ by \cref{lem:non-monotonic kernel}.
		Furthermore, $\valuation_{\vec{e}}(\kernel{npe_\alpha}{\invariant'}{\Delta,\Gamma})$ is non-positive
                for all large enough $n$.
		Hence, it suffices to show that for all $n \geq \max\{ C_\alpha, 2 \cdot
                |\valuation_{\vec{e}}|( \sqcup Pol_\alpha) \}$, the inequation
                $\valuation_{\vec{e}}(\kernel{npe_\alpha}{\invariant'}{\Delta,\Gamma}) \leq 0$ is always
                fulfilled, because this means that $\valuation_{\vec{e}}(npe_\alpha)\leq 0$ holds, too.

		If $\valuation_{\vec{e}}(\kernel{npe_\alpha}{\invariant'}{\Delta,\Gamma}) = 0$, then the
                claim is trivial.

		So from now on let $\valuation_{\vec{e}}(\kernel{npe_\alpha}{\invariant'}{\Delta,\Gamma}) \neq 0$.
		Thus, there exists a maximal index $1 \leq \ell_{\vec{e}} \leq \ell$ where $\valuation_{\vec{e}}(p_{\ell_{\vec{e}}}) \neq 0$.
		If $\ell_{\vec{e}} = 1$, then the sign of
                $\valuation_{\vec{e}}(\kernel{npe_\alpha}{\invariant'}{\Delta,\Gamma}) =
                \valuation_{\vec{e}}(p_1) \cdot n^{a_1}\cdot b_1^n$ is determined by
                $\sign(\valuation_{\vec{e}}(p_1))$ for every $n \geq 1$. (Recall that $p_1
                \in \ZZ[\vec{x}]$.)
		Since $\valuation_{\vec{e}}(\kernel{npe_\alpha}{\invariant'}{\Delta,\Gamma}) \leq 0$ holds
                for large enough $n$, it also holds
                for all $n \geq \max\{ C_\alpha,  2 \cdot |\valuation_{\vec{e}}|( \sqcup Pol_\alpha) \} \geq 1$.

		Otherwise we have $\ell_{\vec{e}} = j$ for some $2 \leq j \leq \ell$.
		Then we have the following for all $n \geq N_j$:
		\begin{align}
			        & \abs*{\sum_{i=1}^{j-1} \valuation_{\vec{e}}(p_i) \cdot n^{a_i} \cdot b_i^n}    \nonumber                                                                                                               \\
			\leq {} & \sum_{i=1}^{j-1}\abs*{ \valuation_{\vec{e}}(p_i)} \cdot n^{a_i} \cdot b_i^n\nonumber                                                                                                          \\
			\leq {} & \sum_{i=1}^{j-1} |\valuation_{\vec{e}}|(\sqcup \{p_1,\ldots,p_{j-1} \}) \cdot n^{a_i} \cdot b_i^n\tag{by \cref{Soundness of Over-approximation}}\nonumber                                \\
			= {}    & |\valuation_{\vec{e}}|(\sqcup \{p_1,\ldots,p_{j-1} \})\cdot \Big(n^{a_{j-1}}\cdot b_{j-1}^n + \sum_{i=1}^{j-2} n^{a_i} \cdot b_i^n\Big)\nonumber                                         \\
			\leq{}  & |\valuation_{\vec{e}}|(\sqcup \{p_1,\ldots,p_{j-1} \})\cdot \Big(n^{a_{j-1}}\cdot b_{j-1}^n + \sum_{i=1}^{j-2} n^{a_{j-2}}\cdot b_{j-2}^n\Big) \tag{as $n \geq N_j \geq mt$ for $j > 3$}
			\\
			={}
			        & |\valuation_{\vec{e}}|(\sqcup \{p_1,\ldots,p_{j-1} \})\cdot \Big(n^{a_{j-1}}\cdot b_{j-1}^n + (j-2)\cdot n^{a_{j-2}}\cdot b_{j-2}^n\Big)\nonumber                                        \\
			\leq{}  & 2 \cdot |\valuation_{\vec{e}}|(\sqcup \{p_1,\ldots,p_{j-1} \})\cdot n^{a_{j-1}}\cdot b_{j-1}^n\tag{as $n \geq N_j \geq mt'$ for $j \geq 3$}
		\end{align}
		Note that if $|\valuation_{\vec{e}}|(\sqcup \{p_1,\ldots,p_{j-1} \}) \neq 0$, then the last inequation is strict.

		Clearly, $(b_j, a_j) \lex (b_{j-1},a_{j-1})$ implies $b_j > b_{j-1}$ or both $b_j = b_{j-1}$ and $a_j \geq a_{j-1} + 1$.
		If $b_j = b_{j-1}$ and $a_j \geq a_{j-1} + 1$, we have that
		\begin{align*}
			2 \cdot |\valuation_{\vec{e}}|(\sqcup \{p_1,\ldots,p_{j-1} \})\cdot n^{a_{j-1}}\cdot b_{j-1}^n & = 2 \cdot |\valuation_{\vec{e}}|(\sqcup \{p_1,\ldots,p_{j-1} \})\cdot n^{a_{j-1}}\cdot b_{j}^n \\
			                                                                                               & \leq n^{a_{j-1} + 1}\cdot b_{j}^n                                                              \\
			                                                                                               & \leq n^{a_j}\cdot b_{j}^n
		\end{align*}
		holds for all $n \geq 2 \cdot |\valuation_{\vec{e}}|(\sqcup \{p_1,\ldots,p_{j-1}
			\})$, where the but-last inequation is strict if $|\valuation_{\vec{e}}|(\sqcup \{p_1,\ldots,p_{j-1}
			\}) = 0$ and $n \geq 1$.

		In the second case $b_j > b_{j-1}$, we can derive $$ 2 \cdot |\valuation_{\vec{e}}|(\sqcup \{p_1,\ldots,p_{j-1} \})\cdot n^{a_{j-1}}\cdot b_{j-1}^n \leq n^{a_{j-1} + 1}\cdot b_{j-1}^n < n^{a_j}\cdot b_{j}^n$$ for all $n \geq \max\{M_j,2 \cdot |\valuation_{\vec{e}}|(\sqcup \{p_1,\ldots,p_{j-1} \})\}$, as $M_j$ is the 1-monotonicity threshold of $(b_j,a_j)$ and $(b_{j-1}, a_{j-1} + 1)$.
		Thus, in total we have shown that for all $n \geq \max\{N_j,M_j,2\cdot|\valuation_{\vec{e}}|(\sqcup \{p_1,\ldots,p_{j-1} \})\}$, we have
		\begin{equation}
			\abs*{\sum_{i=1}^{j-1} \valuation_{\vec{e}}(p_i) \cdot n^{a_i} \cdot b_i^n} < n^{a_j}\cdot b_j^n.\label{proof:eqSTH}
		\end{equation}
		Now we prove that
                $\valuation_{\vec{e}}(\kernel{npe_\alpha}{\invariant'}{\Delta,\Gamma}) <
                0$ holds for all $n \geq \max\{N_j,M_j,2\cdot|\valuation_{\vec{e}}|(\sqcup
                \{p_1,\linebreak  \ldots,p_{j-1} \})\}$.
		As $\valuation_{\vec{e}}(\kernel{npe_\alpha}{\invariant'}{\Delta,\Gamma})$ is non-positive
                for large enough $n$, we must have
                $\valuation_{\vec{e}}(p_j) \leq 0$.
		However, $\valuation_{\vec{e}}(p_j) = 0$ is prevented by the definition of $\ell_{\vec{e}} = j$.
		Thus, we have $\valuation_{\vec{e}}(p_j) < 0$ and hence,
		\begin{align*}
			       & \valuation_{\vec{e}}(\kernel{npe_\alpha}{\invariant'}{\Delta,\Gamma})                                                                                                                                      \\
			={}    & \valuation_{\vec{e}}(p_j)\cdot n^{a_j}\cdot b_j^n + \sum_{i=1}^{j - 1} \valuation_{\vec{e}}(p_i) \cdot n^{a_i} \cdot b_i^n\nonumber                                                                    \\
			\leq{} & \valuation_{\vec{e}}(p_j)\cdot n^{a_j}\cdot b_j^n + \abs*{\sum_{i=1}^{j-1} \valuation_{\vec{e}}(p_i) \cdot n^{a_i} \cdot b_i^n}\tag{as $x + y \leq x + \abs*{y}$ holds for all $x,y \in \ZZ$}\nonumber \\
			<{}    & \valuation_{\vec{e}}(p_j)\cdot n^{a_j}\cdot b_j^n + n^{a_j}\cdot b_j^n\tag{by \eqref{proof:eqSTH}}\nonumber                                                                                       \\
			={}    & (\valuation_{\vec{e}}(p_j) + 1) \cdot n^{a_j}\cdot b_j^n                                                                                                                                          \\
			\leq{} & 0 \tag{since $p_j \in \ZZ[\vec{x}]$ and thus $\valuation_{\vec{e}}(p_j) < 0$ implies $\valuation_{\vec{e}}(p_j) +1 \leq 0$}
		\end{align*}

		Note that we have $C_\alpha \geq N_j$ and $C_\alpha \geq M_j$ for all $2 \leq j \leq \ell$.
		Moreover, since $Pol_\alpha \supseteq \{p_1,\ldots,p_{j-1}
			\}$, we have $|\valuation_{\vec{e}}|(\sqcup Pol_\alpha) \geq |\valuation_{\vec{e}}|(\sqcup \{p_1,\ldots,p_{j-1}
			\})$. Hence, for all $n \geq \max\{C_\alpha, D_\alpha, 2 \cdot |\valuation_{\vec{e}}|(\sqcup Pol_\alpha)\} \geq \max\{N_j,M_j,D_\alpha,2\cdot|\valuation_{\vec{e}}|(\sqcup \{p_1,\ldots,p_{j-1}
			\})\}$, the inequation $\valuation_{\vec{e}}(npe_\alpha) \leq 0$ holds.
		Hence, we have $\max\{ n_0, C_\alpha, D_\alpha, 2 \cdot
                |\valuation_{\vec{e}}|( \sqcup Pol_\alpha) \}\linebreak
                 \geq \; \sth_{(\invariant,a,\update)}(\vec{e})$.

Clearly, we have $|\valuation_{\vec{e}}|(\sthb) \geq \max\{ n_0, C, D, 2 \cdot
|\valuation_{\vec{e}}|( \sqcup Pol) \}$. Hence, it suffices to prove that $\max\{ n_0, C, D, 2 \cdot
|\valuation_{\vec{e}}|( \sqcup Pol) \} \geq \sth_{(\invariant,\guard,\update)}(\vec{e})$
holds.   	The stabilization threshold $\sth_{(\invariant,\guard,\update)}(\vec{e})$
is at most the maximum of the stabilization thresholds
$\sth_{(\invariant,a,\update)}(\vec{e})$ for those atoms $\alpha$ in $\guard$ where
where
$\valuation_{\vec{e}}(\update^n(\alpha)) = \false$ for all large enough $n$.
		Hence, for $\max\{ n_0, C,
		D,	2 \cdot |\valuation_{\vec{e}}|( \sqcup Pol) \} \; \geq \;
                \sth_{(\invariant,\guard,\update)}(\vec{e})$, by \eqref{complexityClaim}
                it suffices to show that $\max\{ n_0, C,D, 2 \cdot |\valuation_{\vec{e}}|(
                \sqcup Pol) \} \geq \max\{ n_0, C_\alpha, D_\alpha, 2 \cdot
                |\valuation_{\vec{e}}|(\sqcup Pol_\alpha)\}$ holds for all these atoms $\alpha$ in $\guard$. This immediately follows from $C \geq C_\alpha$, $D \geq D_\alpha$, and $Pol \supseteq Pol_\alpha$ which implies $|\valuation_{\vec{e}}|( \sqcup Pol) \geq |\valuation_{\vec{e}}|(\sqcup Pol_\alpha)$.

                Note that since $\invariant$ is the guard of $t_0$,
                $\guard$ is the guard of $t$, and $\invariant' = (\invariant \land \guard)$,
                an evaluation of $t$ can only occur in
the program $\{ t_0, t \}$ if the initial configuration has the form
$(\location_0, \valuation_{\vec{e}})$ for some $\vec{e} \in \InvSet'$.
For such a $\valuation_{\vec{e}}$,
recall that for all $n \geq \sth_{(\invariant,\guard,\update)}(\vec{e})$, the truth value of $\valuation_{\vec{e}}(\update^n(\guard))$ remains the same.
		Hence, if the twn-loop is terminating, then we have $\valuation_{\vec{e}}(\update^n(\guard)) = \false$ for all these $n$.
		So for all evaluations $(\location_0, \valuation_{\vec{e}}) \rightarrow_{t_0} (\location, \valuation_{\vec{e}}) \rightarrow_t^k (\location, \valuation_{\vec{e}}')$ with the looping transition $t$, we must have $k \leq \sth_{(\invariant,\guard,\update)}(\vec{e})$.
		Hence, $\glo$ with $\glo(t) = \sthb$ is a global runtime bound, since $|\valuation_{\vec{e}}|(\sthb) \geq \; \sth_{(\invariant,\guard,\update)}(\vec{e})$.
	\end{myproof}
}

\begin{example}
	\label{exa:twnselfloops_global}
The guard $\guard$ of the tnn-loop in \cref{ex:tnn}
 has the atoms $\alpha = (x_1^2 + x_3^5 < x_2)$,
 $\alpha' = (0 < x_1)$, and $\alpha'' = (0 <
 -x_1)$ (since $x_1 \neq 0$ is transformed into $\alpha'  \lor \alpha''$).
	When instantiating the variables by the closed forms of \cref{ex:closed form} with
        start value $n_0 = 0$,
 \Cref{thm:complexity} computes the bound 1 on the stabilization thresholds for $\alpha'$ and\linebreak $\alpha''$.
 So the only interesting atom is $\alpha = (0 < s_2 - s_1)$ for $s_1
 = x_1^2 + x_3^5$ and $s_2 = x_2$.\linebreak
We get
        $npe_\alpha = (s_2 - s_1)[\vec{x}/\cl{\vec{x}}] =
        q_3 \cdot 16^n + q_2 \cdot 9^n + q_1$, with
        $q_j$ as in \Cref{ex:kernel}.

	In the program of \cref{fig:twnselfloop}, the corresponding self-loop
        $t_5$ has two entry transitions $t_4$ and $t_1$ which result in two
   tnn-loops with the  update-invariants
        $\invariant_1 = \true$ resulting from transition $t_4$ and $\invariant_2 = (x_3 > 0)$
        from $t_1$. So $\invariant_2$ is an update-invariant of $t_5$
        which always holds when reaching $t_5$ via transition $t_1$.

	For $\invariant_1 = \true$, we  choose
        $\Delta = \Gamma = \emptyset$, i.e., $\kernel{npe_\alpha}{\invariant'_1}{\Delta,\Gamma}
        = npe_\alpha$. 	So we have  $b_3 = 16$, $b_2 = 9$,  $b_1 = 1$, and $a_j = 0$ for all $1
        \leq j \leq 3$.
	We obtain
	\[
		\begin{array}{rcll}
			M_2 & = & 0, & \text{as 0 is the 1-monotonicity threshold of $(9,0)$ and $(1,1)$}        \\
			M_3 & = & 0, & \text{as 0 is the 1-monotonicity threshold of
                          $(16,0)$ and $(9,1)$}  \\
		N_2 = 1 \; \text{ and }	\; N_3 & = & 1, & \text{as 1 is the 1-monotonicity threshold of $(9,0)$ and $(1,0)$}.        \\
		\end{array}
	\]
	Hence, we get $C = C_\alpha = \max\{1, N_2, M_2, N_3, M_3\} = 1$.
	So we obtain the runtime bound $\sthbInv{\invariant_1} = 2 \cdot \sqcup \{
        q_1, q_2 \} + \max \{n_0, C_\alpha\} = 2 \cdot x_2 + 2\cdot x_3^3 + 2\cdot
        x_3^5 + 1$ for the  loop $t_5 \chain t_5$  w.r.t.\ $\invariant_1$.
By \cref{Chaining Preserves Asymptotic Runtime},
 this means that $2\cdot \sthbInv{\invariant_1} + 1 = 4 \cdot x_2 + 4\cdot x_3^3 + 4\cdot x_3^5 + 3$ is a runtime bound for the loop at transition $t_5$.

For the update-invariant $\invariant_2 = (x_3 > 0)$,
        we use the over-approximation\report{\linebreak}
      $\kernel{npe_\alpha}{\invariant_2'}{\Delta,\Gamma} = p_2 \cdot 16^n + p_1 \cdot 9^n$
        with $p_2 = -x_1^2$ and $p_1 = x_2$ from \Cref{ex:kernel}, where $\invariant_2' =\linebreak
(\invariant_2 \land \guard)$ implies that it is always non-positive
        for large enough $n$.
	Now we obtain
        $M_2  =  0$ (the 1-monotonicity threshold of $(16,0)$ and $(9,1)$) and
        $N_2  =  1$,
        where $C = C_\alpha = \max \{ 1, N_2, M_2 \} = 1$. Moreover, we have
 $D_\alpha = \max \{ 1,0 \} = 1$,
        since
	\[
		\begin{array}{l}
		\text{1 is the 1-monotonicity threshold of $(9,0)$ and $(1,0)$, and } \\
	\text{0 is the 1-monotonicity threshold of $(1,0)$ and $(0,0)$.} \\
                \end{array}
	\]
            We now get the tighter bound $\sthbInv{\invariant_2} = 2 \cdot \sqcup
     \{ p_1 \} + \max \{n_0, C_\alpha, D_\alpha\} = 2\cdot x_2 + 1$ for\linebreak  $t_5\chain t_5$.
	So $t_5$'s runtime bound is $2\cdot \sthbInv{\invariant_2} + 1 = 4\cdot
        x_2 + 3$ when using invariant $\invariant_2$.
\end{example}

\Cref{cor:complexity} shows how  the technique of \cref{Chaining Preserves Asymptotic Runtime,thm:complexity} can be used to compute local runtime bounds for twn-loops whenever such loops occur within an integer program.
To this end, one needs the new \cref{thm:time-bound} where in contrast to \cite{brockschmidt2016AnalyzingRuntimeSize,Festschrift} these local bounds do not have to result from ranking functions.

To turn a self-loop $t$ and  $\pret \in \entry_{\{t\}}$ from a larger
program $\PP$ into a twn-loop $\IntLoop$,
we use $t$'s guard $\guard$ and update $\update$.
To obtain an update-invariant $\invariant$, our implementation
uses the \tool{Apron} library \cite{jeannet2009ApronLibraryNumerical} for computing invariants
on a version of the full program where we remove all entry transitions $\entry_{\{ t \}}$
except $\pret$.\footnote{\label{local restricted 2}Regarding invariants for the full program in the computation
of local bounds for $t$ is
possible since in contrast to \cite{brockschmidt2016AnalyzingRuntimeSize,Festschrift} our
definition of local bounds from \Cref{def:locUpperTimeBound} is
restricted to states that are reachable from an initial configuration $(\location_0, \valuation_0)$.}
From the invariants computed for $t$, we take those that are also
update-invariants of $t$.

\begin{theorem}[Local Bounds for Twn-Loops]
	\label{cor:complexity}
	Let
	$\PP = \IntProgram$ be an integer program with $\PVSet' = \{x_1,\ldots,x_d\}
		\subseteq \PVSet$.
	Let $t = (\location, \varphi, \update, \location)\in\TSet$  with $\guard \in
        \FormulaSet(\PVSet')$, $\update(v) \in \ZZ[\PVSet']$ for all $v \in \PVSet'$, and
        $\update(v) = v$ for all  $v \in \PVSet\setminus\PVSet'$.
	For any entry transition $\pret \in \entry_{\{t\}}$,
let  $\invariant \in \FormulaSet(\PVSet')$
such that
 $\models \invariant \to
\update(\invariant)$
and such that
$\valuation(\invariant)$ holds whenever there is a
$\valuation_0 \in \Valuation$ with $(\location_0, \valuation_0)
\rightarrow_{\TSet}^* \circ  \rightarrow_{\pret} (\location, \valuation)$.
	If $L=\IntLoop$ is a terminating tnn-loop, then let $\loc(\rightarrow_\pret \braced{t}) = \sthb$,
	where $\sthb$ is defined as in \cref{thm:complexity}.
	If $L$ is a terminating twn-loop but no tnn-loop, let $\loc(\rightarrow_\pret \braced{t}) = 2 \cdot \sthb + 1$, where $\sthb$ is the bound of \cref{thm:complexity}
	computed for $L \chain L$.
        Otherwise, let $\loc(\rightarrow_\pret \braced{t}) = \omega$.
	Then $\loc$ is a local runtime bound for $\braced{t} = \TSet'_>= \TSet'$ in the
        program $\PP$.
        \end{theorem}
 \makeproof{cor:complexity}{
	\begin{myproof}
		We want to prove that $\loc : \entry_{\braced{t}}\to\BoundSet$  is a local
                runtime bound according to \cref{def:locUpperTimeBound}.
		Let $\pret\in\entry_{\braced{t}}$ be an entry transition.
	        We only prove the case where $\IntLoop$ is a terminating tnn-loop (the
                case of a terminating twn-loop that is not a tnn-loop then follows by
                \cref{Chaining Preserves Asymptotic Runtime} and the case where
$\IntLoop$ is not a terminating twn-loop is trivial).
		By \cref{thm:complexity},
when starting in a configuration $(\location, \valuation_{\vec{e}})$ for $\vec{e} \in
\InvSet'$, the transition $t$ can be  evaluated at most $\sthb(\vec{e})$
times.
		By construction of $\invariant$, all states $\valuation$  satisfy
                $\invariant' = (\invariant \land \guard)$ if they result from
                an evaluation step with $\pret$
before a next step with $t$ in an execution of the full program.
	Thus, for all $\valuation\in\Valuation$, we have $$
\begin{array}{rcl}
        |\valuation|(\sthb) &\geq&
  \sup \braced{ k \in \NN \mid \exists\, \initial,
                \valuation'. \;
(\location_0, \initial) \rightarrow_{\TSet}^* \circ  \rightarrow_{\pret} \, (\location, \valuation) \; \rightarrow_t^k \; (\location, \valuation') }.
\end{array}
        $$
	\end{myproof}
 }

\begin{example}
	\label{ex:twnloopInLargerProgram}
	In \cref{fig:twnselfloop}, we consider the self-loop $t_5$ with
$\entry_{\{t_5\}} = \{ t_4, t_1 \}$
       and the update-invariants $\invariant_1 = \true$
        resp.\ $\invariant_2 = (x_3 > 0)$.
	For $t_5$'s guard $\guard$
        and update $\update$,
        both
        $(\invariant_i, \guard, \update)$ are terminating
        twn-loops (see \cref{ex:twn}), i.e., \eqref{fml ev non-term} is invalid.

	By \cref{cor:complexity,exa:twnselfloops_global}, $\loc$ with
$\loc(\rightarrow_{t_4} \braced{t_5}) = 4\cdot x_2 + 4\cdot x_3^3 + 4\cdot x_3^5 + 3$ and
        $\loc(\rightarrow_{t_1} \braced{t_5}) = 4\cdot x_2 + 3$
        is a local runtime bound for $\{t_5\}= \TSet'_>= \TSet'$ in the program of \cref{fig:twnselfloop}.
	As shown in \cref{ex:fullExample}, \cref{thm:time-bound}
	then yields the global runtime bound $\glo(t_5) = 8 \cdot x_4 \cdot x_5 +
        13006\cdot x_4$.
\end{example}

\section{Local Runtime Bounds for Twn-Cycles}
\label{Local Runtime Bounds for TWN-Cycles}

\Cref{Local Runtime Bounds for TWN-Self-Loops} introduced a technique to determine local runtime bounds for twn-self-loops in a program.
To increase its applicability, we now extend it to larger cycles.
For every entry transition of the cycle,
we \emph{chain} the transitions of the cycle, starting with the transition which follows the entry transition.
In this way, we obtain loops  consisting of a single transition.
 If the chained loop is a twn-loop, we can apply \cref{cor:complexity} to compute a local
runtime bound.
\pagebreak[2] Any
 local bound on the chained transition
 is also a bound on each of the original transitions.\footnote{This is sufficient for our
 improved definition of local bounds in \Cref{def:locUpperTimeBound} where in contrast to
 \cite{brockschmidt2016AnalyzingRuntimeSize,Festschrift} we do not
 require a bound on the \emph{sum} but only on \emph{each}
transition in the considered  set $\TSet'$.
Moreover, here we again benefit from our extension to compute individual local
bounds for different entry transitions.}

By \cref{cor:complexity}, we obtain a bound on the number of evaluations of the
\emphit{complete cycle}.
However, we also have to consider a \emph{partial execution} which stops before traversing
the full cycle. Therefore,
we increase every local runtime bound by 1.

Note that this replacement of a cycle by a self-loop which results from chaining its
transitions is only sound
for \emph{simple} cycles.
A cycle is simple if each iteration through the cycle can only be done in a unique way.
So the cycle must not have any subcycles and there also must not be any indeterminisms
concerning the next transition to be taken.
Formally, $\mathcal{C}= \{t_1,\ldots,t_n\}\subset\TSet$ is a simple cycle if
$\mathcal{C}$ does not contain temporary variables and there are pairwise different
locations $\location_1,\ldots,\location_n$ such that
$t_i = (\location_i, \_, \_,
\location_{i+1})$ for $1 \leq i \leq n-1$
and $t_n = (\location_n, \_, \_, \location_1)$.
This ensures that if there is an evaluation with $\to_{t_i} \circ \to^*_{\mathcal{C}\setminus\{t_i\}} \circ \to_{t_i}$, then the steps with $\to^*_{\mathcal{C}\setminus\{t_i\}}$ have the form $\to_{t_{i+1}} \circ \ldots \circ \to_{t_n} \circ \to_{t_1} \circ \ldots \circ \to_{t_{i-1}}$.

\begin{figure}[t]
\begin{algorithm}[H]
	\DontPrintSemicolon
	\caption{Algorithm to Compute Local Runtime Bounds for Cycles}\label{alg:twncycleSym}
	\Input{A program $\IntProgram$ and a simple cycle $\mathcal{C}=
	  \{t_1,\ldots,t_n\}\subset\TSet$}
  	\Output{A local runtime bound $\loc$ for $\mathcal{C}= \TSet'_>= \TSet'$}
	Initialize $\loc$: $\loc(\rightarrow_{\pret} \mathcal{C} ) = \omega$ for all
	$\pret\in\entry_{\mathcal{C}}$. \label{alg:initial}\\
	\ForAll{$\pret \in\entry_{\mathcal{C}}$ \label{alg:forloop}} {
		Let $i \in \{1,\ldots,n\}$ such that
		$\pret$'s target location is the start location $\location_i$ of $t_i$. \label{alg:startLocation}\\
		Let $t = t_i \chain \ldots \chain t_n \chain t_1 \chain \ldots \chain
			t_{i-1}$. \label{alg:chaining}\\
		\If{there exists a renaming $\pi$ of $\PVSet$
			such that
			$\pi(t)$ results in a twn-loop
			\label{alg:if}
		}
		{		Set
			$\loc(\rightarrow_{\pret}\mathcal{C}) \gets \text{$\pi^{-1}(1 \, + $ result of
						\Cref{cor:complexity} on $\pi(t)$ and $\pi(r))$.}$
			\label{alg:then} 		}
	}
	\return local runtime bound $\loc$.
\end{algorithm}
\vspace*{-.4cm}
\end{figure}

\Cref{alg:twncycleSym} describes how to compute a local runtime bound for a simple cycle
$\mathcal{C} = \{ t_1, \ldots, t_n \}$ as above.
In the loop of \Cref{alg:forloop}, we iterate over all entry transitions\linebreak $\pret$ of $\mathcal{C}$.
If $\pret$ reaches the transition $t_i$, then in
\Cref{alg:startLocation,alg:chaining} we chain $t_i \chain \ldots \chain t_n \chain t_1
\chain\linebreak \ldots \chain t_{i-1}$ which corresponds to one iteration of the cycle starting in $t_i$.
If a suitable renaming (and thus also reordering) of the variables turns the chained transition into a twn-loop, then we use \cref{cor:complexity} to compute a local runtime bound $\loc(\rightarrow_{\pret}\mathcal{C})$ in \Cref{alg:if,alg:then}.
If the chained transition does not give rise to a twn-loop, then
$\loc(\rightarrow_{\pret}\mathcal{C})$ is $\omega$ (\Cref{alg:initial}).
In practice, to use the twn-technique for
a transition $t$ in a program,
our tool \KoAT{} searches for those
simple cycles that contain $t$ and  where the chained cycle is a twn-loop. Among those
cycles it chooses the one with the smallest
runtime bounds for its entry transitions.
\begin{theorem}[Correctness of \Cref{alg:twncycleSym}]
	\label{lem:twncycleSym}
	Let $\PP = \IntProgram$ be an integer program and
	let $\mathcal{C}\subset\TSet$ be a simple cycle in $\PP$.
	\pagebreak[2] Then the result $\loc : \entry_{\mathcal{C}} \rightarrow \BoundSet$ of
        \Cref{alg:twncycleSym} is a local runtime bound for $\mathcal{C} = \TSet'_> = \TSet'$.
\end{theorem}
\makeproof{lem:twncycleSym}{
	\begin{myproof}
		Let $\pret\in\entry_{\mathcal{C}}$.
		If $\loc(\rightarrow_{\pret}\mathcal{C}) = \omega$, then the claim is trivial.
		Otherwise, let $\location_i$ be the target location of $\pret$, let $1 \leq j \leq n$, and let $\valuation\in\Valuation$.
		By \cref{cor:complexity} which yields a local runtime bound for $\{ \pi(t) \}
			= \pi(\TSet_{>}') = \pi(\TSet')$ and hence also for $\{ t \}
			= \TSet_{>}' = \TSet'$, we obtain
		\[
			\begin{array}{cl}
				     & |\valuation|(\loc(\rightarrow_{\pret}\mathcal{C}))                                                                                                                                                                                                                 \\
			  \geq & 1 + \sup \{ k \in \NN \mid
\exists\, \initial, \valuation'. \;
(\location_0, \initial) \rightarrow_{\TSet}^* \circ
 \rightarrow_{\pret}
				(\location_i, \valuation) \to^k_{t_i \chain \ldots \chain t_n \chain t_1 \chain \ldots \chain t_{i-1}}
				(\location_i, \valuation') \}                                                                                \\
 =    & \sup \{ k +1 \mid
\exists\, \initial, \valuation'. \;                                                        \\
&\;(\location_0, \initial) \rightarrow_{\TSet}^* \circ
 \rightarrow_{\pret} \; (\location_i, \valuation) \; (\to_{t_i} \circ ... \circ \to_{t_{j-1}} \circ \to_{t_j} \circ \to_{t_{j+1}} \circ ... \to_{t_{i-1}})^k \; (\location_i, \valuation') \}                              \\
 \geq & \sup \{ k \in \NN \mid \exists\,
 \initial, (\location',\valuation'). \;
(\location_0, \initial) \rightarrow_{\TSet}^* \circ \rightarrow_{\pret} \; (\location_i, \valuation) \; (\to^*_{\mathcal{C}\setminus\{t_j\}} \circ \to_{t_j})^k \; (\location', \valuation') \},
			\end{array}
		\]
		since $\mathcal{C}$ is a simple cycle.
	\end{myproof}
}
\begin{example}
	We apply \Cref{alg:twncycleSym} on the cycle $\mathcal{C} = \braced{t_{5a},t_{5b}}$ of the program in \Cref{fig:twncycle}.
	$\mathcal{C}$'s entry transitions $t_1$ and $t_4$ both end in $\location_3$.
	Chaining $t_{5a}$ and $t_{5b}$ yields the transition $t_5$ of \cref{fig:twnselfloop}, i.e., $t_5 = t_{5a}
		\chain t_{5b}$. Thus, \Cref{alg:twncycleSym} essentially transforms the
                program of \Cref{fig:twncycle} into \cref{fig:twnselfloop}. As in
                \cref{exa:twnselfloops_global,ex:twnloopInLargerProgram},
                we obtain
$\loc(\rightarrow_{t_4}\mathcal{C}) = 1 + (2\cdot\sthbInv{\true} + 1) = 4 \cdot x_2 +
                4\cdot x_3^3 + 4\cdot x_3^5 + 4$ and
                $\loc(\rightarrow_{t_1}\mathcal{C}) = 1 + (2\cdot\sthbInv{x_3 > 0} + 1) =
                4\cdot x_2 + 4$, resulting in
the global runtime bound $\glo(t_{5a}) = \glo(t_{5b}) = 8
                 \cdot x_4 \cdot x_5 + 13008 \cdot x_4$, which again yields $\rc(\initial) \in \landau(n^2)$.
\end{example}

\paper{\begin{figure}[t]}\report{\begin{figure}[h]}
	\centering
	\hspace*{-.6cm}	\begin{tikzpicture}[->,>=stealth',shorten >=1pt,auto,node distance=3.5cm,semithick,initial text=$ $]
		\node[state,initial] (q0) {$\location_0$};
		\node[state] (q1) [right of=q0,xshift=-2cm]{$\location_1$};
		\node[state,draw=none] (h0) [right of=q1, node distance=2cm]{};
		\node[state] (q2) [below of=h0, node distance=1.2cm]{$\location_2$};
		\node[state] (q3) [right of=q1, node distance=4cm]{$\location_3$};
		\node[state] (q4) [right of=q3, node distance=5.2cm]{$\location_4$};
		\draw (q0) edge node [text width=1.5cm,align=center] {{\scriptsize
                            $t_0$}} (q1);
		\draw (q1) edge[bend left=80, looseness=.45] node [text
                  width=4cm,align=center,below,yshift=-.14cm] {{\scriptsize $t_1:\guard =
                    x_3 > 0 \wedge x_4 > 0$ \\
	    $\begin{array}{rcl}
                              \update(x_1) &=& x_4 \\
			      \update(x_2) &=& x_5
\end{array}$}} (q3);
		\draw (q1) edge[bend right] node [text width=4cm,align=center,below,
                  xshift=-1.8cm, yshift=.2cm] {{\scriptsize $t_3:\guard =	-5 \leq
                    x_3 \leq 5\;$}} (q2);
  	\draw (q2) edge[bend right] node [text width=4cm,align=center,below,  xshift=1.57cm, yshift=.3cm] {{\scriptsize $t_4:\guard = x_4 > 0$ \\
			  $\begin{array}{rcl}
		            \update(x_1) &=& x_4 \\
			    \update(x_2) &=& x_5
            \end{array}$}} (q3);
              	\draw (q3) edge[bend left=75, looseness=.35] node [text
                  width=5cm,align=center,below,yshift=-.08cm] {{\scriptsize $t_{5a}:\guard = x_1^2 + x_3^5 < x_2 \wedge x_1 \neq 0$ \\
		 \centerline{$\update(x_1) = -2\cdot x_1$}}} (q4);
		\draw (q4) edge node [text width=3.5cm,align=center,yshift=.05cm]
                      {{\scriptsize $t_{5b}:\update(x_2) = 3\cdot x_2 - 2\cdot x_3^3$}} (q3);
		\draw (q3) edge[bend left=10] node [text width=3cm,align=center] {{\scriptsize $t_2: \update(x_4) = x_4 - 1$}} (q1);
	\end{tikzpicture}
	\caption{\vspace*{-.2cm}An Integer Program with a Nested Non-Self-Loop}\label{fig:twncycle}
\end{figure}

\section{Conclusion and Evaluation}
\label{Evaluation}

We showed that results on subclasses of programs with computable complexity bounds like
\cite{hark2020PolynomialLoopsTermination} are not only theoretically interesting, but they have an important practical value.
To our knowledge, our paper is the first to integrate such results into an
incomplete approach for automated complexity analysis like
\cite{brockschmidt2016AnalyzingRuntimeSize,Festschrift}. \pagebreak[2]
For this integration, we developed several novel contributions which extend and improve
the previous approaches in
\cite{hark2020PolynomialLoopsTermination,brockschmidt2016AnalyzingRuntimeSize,Festschrift}
substantially:
\begin{enumerate}
  \item[(a)] We extended the concept of local runtime bounds such that they can now depend on entry
    transitions (\cref{def:locUpperTimeBound}).
\item[(b)] We generalized the computation of global runtime bounds such that one can now lift
    arbitrary local bounds to global bounds (\cref{thm:time-bound}). In particular,
    the local bounds might be due to either ranking functions or twn-loops.
\item[(c)] We improved the technique for the computation of bounds on twn-loops such that these
    bounds now take the roles of the different variables into account
    (\Cref{Over-Approximation of Polynomials,Soundness of Over-approximation,thm:complexity}).
\item[(d)] We extended the notion of twn-loops by update-invariants and developed a new
    over-approximation of their closed forms
    which takes invariants into account
    (\Cref{def:twn-loop,def:non-monotonic kernel,lem:non-monotonic kernel,thm:complexity}).
\item[(e)] We extended the handling of twn-loops to twn-cycles (\Cref{lem:twncycleSym}).
\end{enumerate}
      The need for these improvements is
    demonstrated by our leading example in \cref{fig:twnselfloop}\linebreak (where
the contributions (a) - (d) are needed  to infer quadratic runtime complexity) and by  the
example in \cref{fig:twncycle} (which illustrates
(e)). In this way, the power of automated complexity analysis is increased substantially,
because now one can also infer runtime bounds for programs containing non-linear arithmetic.

To demonstrate the power of our approach,
we evaluated the integration of our new technique to infer local runtime bounds for
twn-cycles in our re-implementation of the tool \tool{KoAT} (written in \tool{OCaml})
 and compare the results to other state-of-the-art tools.
To distinguish our re-implementation of \tool{KoAT} from the original version of the tool from \cite{brockschmidt2016AnalyzingRuntimeSize}, let \tool{KoAT1} refer to the tool from \cite{brockschmidt2016AnalyzingRuntimeSize} and let \tool{KoAT2} refer to our new re-implementation.
\tool{KoAT2} applies a local control-flow refinement
technique \cite{Festschrift}  (using the tool \tool{iRankFinder} \cite{domenech2018IRankFinder})
and  preprocesses the program in the beginning, e.g.,
by extending the guards of transitions by invariants inferred using the
\tool{Apron} library
\cite{jeannet2009ApronLibraryNumerical}.
For all occurring SMT problems, \tool{KoAT2} uses \tool{Z3} \cite{moura2008}.
We tested the following configurations of
\tool{KoAT2}, which differ in the techniques used for the computation of local runtime bounds:

\medskip
\hspace*{-.75cm}\begin{minipage}{12.5cm}\begin{itemize}
	\item[$\bullet\!$] \tool{KoAT2\,\!+\,\!RF} only uses linear ranking functions to
          compute local runtime bounds
  	\item[$\bullet\!$] \tool{KoAT2\,\!+\,\!$\MRF 5$} uses
          multiphase-linear ranking functions    of depth $\leq 5$
	\item[$\bullet\!$] \tool{KoAT2\,\!+\,\!TWN} only uses twn-cycles to compute local
          runtime bounds (\Cref{alg:twncycleSym}) \allowbreak
	\item[$\bullet\!$] \tool{KoAT2\,\!+\,\!TWN\,\!+\,\!RF} uses \Cref{alg:twncycleSym} for  twn-cycles and linear ranking functions
	\item[$\bullet\!$] \tool{KoAT2\,\!+\,\!TWN\,\!+\,\!$\MRF 5$}  uses \Cref{alg:twncycleSym}
          for twn-cycles and $\MRFs$ of  depth $\leq 5$
\end{itemize}\end{minipage}

\medskip

Existing approaches  for automated complexity analysis are already very powerful
on programs that only use linear arithmetic in their guards and updates. The\linebreak corresponding
benchmarks for \emph{Complexity of
Integer Transitions Systems}  (\tool{CITS}) and \emph{Complexity of}
\tool{C} \emph{Integer Programs} (\tool{CINT})
from the \emph{Termination Problems Data Base} \cite{tpdb} which is used in the annual \emph{Termination and Complexity Competition (TermComp)} \cite{giesl2019TerminationComplexityCompetition}
contain almost only examples with linear
arithmetic. Here, the
existing tools already infer finite runtimes for more than 89~\% of
those examples in the collections
\tool{CITS} and \tool{CINT}
where this
\emph{might}\footnote{The tool \tool{LoAT}
\cite{frohn2019ProvingNonTerminationLoop,frohn2020InferringLowerRuntime} proves
unbounded runtime for 217 of the 781 examples from
\tool{CITS} and  \tool{iRankFinder} \cite{ben-amram2019MultiphaseLinearRankingFunctions,domenech2018IRankFinder} proves non-termination
for 118
of 484 programs of \tool{CINT}.}
be possible.

The main benefit of our new integration of the twn-technique is that in this way one can
also infer finite runtime bounds for programs that contain non-linear guards or
updates. To demonstrate this, we extended both collections \tool{CITS} and \tool{CINT}
by 20 examples that represent typical such programs, including several
benchmarks from the literature \cite{ben-amram2017MultiphaseLinearRankingFunctions,heizmann2015RankingTemplatesLinear,xu13,Festschrift,frohn2019TerminationTriangularInteger,frohn2020TerminationPolynomialLoops},
as well as our programs from \Cref{fig:twnselfloop,fig:twncycle}.
See \cite{website} for a detailed list and description of these examples.

\begin{figure}[t]
	\makebox[\textwidth][c]{
		\begin{tabular}{l|cc|cc|cc|cc|cc|cc||c|c}
			                                     &
                  \multicolumn{2}{c|}{$\landau(1)$} & \multicolumn{2}{c|}{$\landau(n)$} &
                  \multicolumn{2}{c|}{$\landau(n^2)$} &
                  \multicolumn{2}{c|}{$\landau(n^{>2})$} &
                  \multicolumn{2}{c|}{{\scriptsize $\landau(\mathit{EXP})$}} & \multicolumn{2}{c||}{$< \infty$} & $\mathrm{AVG^+(s)}$ & $\mathrm{AVG(s)}$                                          \\
			\hline {\scriptsize \tool{KoAT2 + TWN + $\MRF 5$}} & 26    &       & 231     & (5)        & 73               & (5)  & 13    & (4) & 1 & (1) & 344 & (15) & 8.72 & 23.93\\
			\hline \tool{KoAT2 + TWN + RF}       & 27    &       & 227     & (5)        & 73               & (5)  & 13    & (4) & 1 & (1) & 341 & (15) & 8.11 & 19.77\\
			\hline \tool{KoAT2 + $\MRF 5$}       & 24    &       & 226     & (1)        & 68               &      & 10   &     & 0 &     & 328 & (1)  & 8.23 & 21.63\\
			\hline \tool{KoAT2 + RF}             & 25    &       & 214     & (1)        & 68               &      & 10   &     & 1 &     & 318 & (1)  & 8.49 & 16.56\\
			\hline \tool{MaxCore}                & 23    &       & 216     & (2)        & 66               &      & 7    &     & 0 &     & 312 & (2)  & 2.02 & 5.31\\
			\hline \tool{CoFloCo}                & 22    &       & 196     & (1)        & 66               &      & 5    &     & 0 &     & 289 & (1)  & 0.62 & 2.66\\
			\hline \tool{KoAT1}                  & 25    &       & 169     & (1)        & 74               &      & 12   &     & 6 &     & 286 & (1)  & 1.77 & 2.77 \\
			\hline \tool{Loopus}                 & 17    &       & 170     & (1)        & 49               &      & 5    & (1) & 0 &     & 241 & (2)  & 0.42 & 0.43\\
			\hline \tool{KoAT2 + TWN}            & 20    &  (1)  & 111     & (4)        & 3                & (2)  & 2    & (2) & 0 &     & 136 & (9)  & 2.54 & 26.59
		\end{tabular}
	}
	\caption{Evaluation on the Collection \tool{CINT${}^+$} \vspace*{-.4cm}}
	\label{fig:CINT}
\end{figure}

\Cref{fig:CINT} presents our evaluation on the collection \tool{CINT${}^+$},
consisting of the  484 examples from \tool{CINT} and our 20 additional
examples for non-linear arithmetic.
We refer to \cite{website} for
the (similar) results on the corresponding collection
\tool{CITS${}^+$}.

In the \tool{C}  programs of \tool{CINT${}^+$}, all
variables are interpreted as integers over $\ZZ$ (i.e., without overflows).
For  \tool{KoAT2} and \tool{KoAT1},
we used \tool{Clang} \cite{clang} and \tool{llvm2kittel} \cite{falke2011TerminationAnalysisPrograms} to transform  \tool{C}
programs into integer transitions systems as in
\Cref{Integer Program}.
We compare \tool{KoAT2} with \tool{KoAT1} \cite{brockschmidt2016AnalyzingRuntimeSize} and the tools \tool{CoFloCo} \cite{flores-montoya2014ResourceAnalysisComplex,cofloco2},
\tool{MaxCore} \cite{albert2019ResourceAnalysisDriven} with \tool{CoFloCo} in the backend, and \tool{Loopus} \cite{sinn2017ComplexityResourceBound}.
We do not compare with \tool{RaML} \cite{ramlpopl17}, as it does not support programs whose complexity depends on (possibly negative) integers (see \cite{ramlweb}).
We also do not compare with \tool{PUBS} \cite{albert2008AutomaticInferenceUpper}, because
as stated in \cite{domenech2019ControlFlowRefinementPartial} by one of its authors, \tool{CoFloCo}
is stronger than \tool{PUBS}.
For the same reason,  we only consider \tool{MaxCore} with the backend \tool{CoFloCo}
instead of  \tool{PUBS}.

All tools were run inside an Ubuntu Docker container on a machine with an AMD Ryzen 7 3700X octa-core CPU and $48 \, \mathrm{GB}$ of RAM.
 As in \emph{TermComp}, we applied a timeout of 5 minutes for every program.

In \Cref{fig:CINT},
the first entry in every cell denotes the number of benchmarks from
\tool{CINT${}^+$} where the respective tool inferred the corresponding bound. The number in
brackets is the corresponding number of benchmarks when only regarding our 20 new examples
for non-linear arithmetic.
The runtime bounds computed by the tools are compared asymptotically as functions which
depend on the largest initial absolute value $n$ of all program variables.
So for instance, there are $26 + 231 = 257$ programs in \tool{CINT${}^+$}
(and 5 of them come from our new examples)
where \tool{KoAT2\,\!+\,\!TWN\,\!+\,\!$\MRF 5$}
can show that $\rc(\valuation_0) \in \landau(n)$ holds for all initial states $\valuation_0$ where $\abs{\valuation_0(v)} \leq n$ for all $v \in \PVSet$.
For $26$ of these programs, \tool{KoAT2\,\!+\,\!TWN\,\!+\,\!$\MRF 5$}
can even show that $\rc(\valuation_0) \in \landau(1)$, i.e., their runtime complexity is constant.
Overall, this configuration succeeds on $344$ examples, i.e., ``$< \infty$'' is the number of examples where a finite bound on the runtime complexity could be computed by the respective tool within the time limit.
``$\mathrm{AVG^+(s)}$'' is the average runtime of the tool on successful runs in seconds, i.e., where the tool inferred a finite time bound before reaching the timeout, whereas ``$\mathrm{AVG(s)}$'' is the average runtime of the tool on all runs including timeouts.

On the original benchmarks \tool{CINT} where
very few examples contain
non-linear arithmetic,
 integrating
\tool{TWN} into
a configuration that already uses multiphase-linear ranking functions
does not increase power much: \tool{KoAT2\,\!+\,\!TWN\,\!+\,\!$\MRF 5$} succeeds on $344-15=329$ such programs and
\tool{KoAT2\,\!+\,\!$\MRF 5$} solves $328-1=327$ examples.
On the other hand, if one only has linear ranking functions,
then an improvement via our twn-technique has similar effects as an improvement with
multiphase-linear ranking functions (here,
the success rate of \tool{KoAT2\,\!+\,\!$\MRF 5$} is similar to \tool{KoAT2\,\!+\,\!TWN\,\!+\,\!RF} which
solves
$341-15=326$ such
programs).

But the main benefit of our technique is that it also allows to successfully handle
examples with non-linear arithmetic. Here, our new technique is significantly
more powerful than previous ones. Other tools and configurations without \tool{TWN} in \Cref{fig:CINT}
solve at most 2 of the
20 new examples.
In contrast, \tool{KoAT2\,\!+\,\!TWN\,\!+\,\!RF} and
\tool{KoAT2\,\!+\,\!TWN\,\!+\,\!$\MRF 5$} both succeed on $15$ of them.\footnote{One
 is the non-terminating
leading example of \cite{frohn2020TerminationPolynomialLoops}, so at most
19 \emph{might} terminate.} In particular,
our running examples from \Cref{fig:twnselfloop,fig:twncycle} and even isolated twn-loops like
$t_5$ or $t_5\chain t_5$ from \Cref{ex:twn,ex:tnn} can \emph{only} be solved by
\tool{KoAT2} with our twn-technique.

To summarize, our evaluations show that \tool{KoAT2} with the added twn-technique
outperforms all other  configurations and tools for automated complexity analysis
on all considered benchmark sets (i.e., \tool{CINT${}^+$}, \tool{CINT}, \tool{CITS${}^+$}, and
\tool{CITS}) and it is the only tool which is also powerful on examples with non-linear arithmetic.

\KoAT's source code, a binary, and a Docker image are available at\paper{ \url{https://aprove-developers.github.io/KoAT_TWN/}.}\report{
\[\mbox{\url{https://aprove-developers.github.io/KoAT_TWN/}.}\]}
The website also has details on our experiments and \emph{web interfaces} to run
\KoAT's   configurations directly online.

\paragraph{Acknowledgments}
We are indebted to M.\ Hark for many fruitful discussions about complexity, twn-loops, and \tool{KoAT}.
We are grateful to S.\ Genaim and J.~J.\ Dom\'enech for  a suitable version of
\tool{iRankFinder}  which we could use for control-flow refinement in \KoAT's backend.
Moreover, we thank A.\ Rubio and E.\ Mart\'in-Mart\'in for  a static binary of
\tool{MaxCore},  A.\ Flores-Montoya and F.\ Zuleger for help in running \tool{CoFloCo} and
\tool{Loopus}, F.\ Frohn for help and advice, and the reviewers for their feedback to
improve the paper.

\paper{\printbibliography}
\report{\bibliographystyle{splncs04}
  \newcommand{\noopsort}[1]{}

}
\end{document}